\documentclass{SciPost}

\hypersetup{
 colorlinks,
 linkcolor={red!50!black},
 citecolor={blue!50!black},
 urlcolor={blue!80!black}
}

\usepackage[bitstream-charter]{mathdesign}
\DeclareSymbolFont{usualmathcal}{OMS}{cmsy}{m}{n}
\DeclareSymbolFontAlphabet{\mathcal}{usualmathcal}

\fancypagestyle{SPstyle}{
\fancyhf{}
\lhead{\colorbox{scipostblue}{\bf \color{white} ~SciPost Physics }}
\rhead{{\bf \color{scipostdeepblue} ~Submission }}

\fancyfoot[C]{\textbf{\thepage}}
}

\usepackage{graphicx}
\usepackage{tabularx}
\usepackage{subcaption}
\usepackage{placeins}
\usepackage{multirow}
\usepackage{graphbox}
\usepackage{xcolor}
\usepackage{authblk}
\usepackage{tikz-feynman}
\tikzfeynmanset{compat=1.0.0}
\usepackage{hyperref}
\usepackage[numbers, sort&compress]{natbib}

\def\beq{\begin{equation}}
\def\eeq{\end{equation}}

\begin{document}

\pagestyle{SPstyle}

\begin{center}{\Large \textbf{\color{scipostdeepblue}{
Constraining the Higgs potential using multi-Higgs production
}}}\end{center}

\begin{center}\textbf{
{Jia-Le~Ding}\textsuperscript{1},
{Zach~Gillis}\textsuperscript{2},
{Ulrich~Haisch}\textsuperscript{3$\star$},
{Brian~Moser}\textsuperscript{4},
{Hai~Tao~Li}\textsuperscript{1},
{Davide~Pagani}\textsuperscript{5$\dagger$},
{Luca~Rottoli}\textsuperscript{6},
{Ambresh~Shivaji}\textsuperscript{7},
{Zong-Guo~Si}\textsuperscript{1},
{Jian~Wang}\textsuperscript{1$\ddagger$},
{Philipp~Windischhofer}\textsuperscript{2,8},
{Xiao~Zhang}\textsuperscript{1}, and 
{Dan~Zhao}\textsuperscript{1}
}\end{center}

\begin{center}
{\bf 1 }{School of Physics, Shandong University, Jinan, Shandong 250100, China}\\
{\bf 2 }{Department of Physics and Enrico Fermi Institute, University of Chicago, Chicago,~IL~60637,~USA}\\
{\bf 3 }{Max Planck Institute for Physics, Boltzmannstr.8, 85748 Garching, Germany}\\
{\bf 4 }{Physikalisches Institut, Universit\"at Freiburg, Hermann-Herder Str. 3a, 79104~Freiburg,~Germany}\\
{\bf 5 }{INFN, Sezione di Bologna, Via Irnerio 46, Bologna, I-40126, Italy}\\
{\bf 6 }{Dipartimento di Fisica G. Occhialini, Università degli Studi di Milano-Bicocca and INFN,~Sezione di Milano-Bicocca, Piazza della Scienza, 3, 20126 Milano, Italy}\\
{\bf 7 }{Indian Institute of Science Education and Research, Knowledge City, Sector 81, S.~A.~S.~Nagar, Manauli PO 140306,
Punjab, India}\\
{\bf 8 }{Kavli Institute for Cosmological Physics, University of Chicago,
Chicago, IL 60637, USA}
\\[\baselineskip]
$\star$ \href{mailto:haisch@mpp.mpg.de}{\small haisch@mpp.mpg.de}\,,\quad
$\dagger$ \href{mailto:davide.pagani@bo.infn.it}{\small davide.pagani@bo.infn.it} \quad
$\ddagger$ \href{mailto:j.wang@sdu.edu.cn}{\small j.wang@sdu.edu.cn}
\end{center}

\begin{center}
 Preprint: LHCHWG-2025-015, MPP-2025-231
\end{center}

\section*{\color{scipostdeepblue}{Abstract}}
\textbf{\boldmath{%
The Higgs self-couplings remain only weakly constrained by current Large Hadron Collider~(LHC) measurements, leaving ample room for physics beyond the Standard Model that could modify the structure of the Higgs potential. Multi-Higgs production processes provide a particularly sensitive probe of deviations in both the Higgs trilinear and quartic self-couplings. In this note, we summarize the current status of next-to-leading-order electroweak~(EW)~corrections to double-Higgs production computed within the Standard Model Effective Field Theory and Higgs Effective Field Theory frameworks, emphasizing how these calculations introduce sensitivity to the Higgs self-couplings beyond what is accessible at leading order. We discuss the key conceptual and technical differences between the two effective field theory approaches, including their treatment of higher-dimensional operators, renormalization procedures, and the structure of EW~two-loop amplitudes. Despite these differences, both approaches yield broadly consistent constraints, illustrating the complementarity of double- and triple-Higgs measurements. With the high-luminosity LHC and future high-energy colliders on the horizon, these developments and further advances provide an essential foundation for extracting increasingly precise information on the dynamics of EW~symmetry breaking.
}}

\vspace{\baselineskip}

\noindent\textcolor{white!90!black}{%
\fbox{\parbox{0.975\linewidth}{%
\textcolor{white!40!black}{\begin{tabular}{lr}%
 \begin{minipage}{0.6\textwidth}%
 {\small Copyright attribution to authors. \newline
 This work is a submission to SciPost Physics. \newline
 License information to appear upon publication. \newline
 Publication information to appear upon publication.}
 \end{minipage} & \begin{minipage}{0.4\textwidth}
 {\small Received Date \newline Accepted Date \newline Published Date}%
 \end{minipage}
\end{tabular}}
}}
}



\newpage

\vspace{10pt}
\noindent\rule{\textwidth}{1pt}
\tableofcontents
\noindent\rule{\textwidth}{1pt}
\vspace{10pt}

\section{Introduction}
\label{sec:introduction}

The discovery of the Higgs boson~\cite{ATLAS:2012yve,CMS:2012qbp} stands as a major milestone in particle physics. Achieving precise measurements of its couplings remains a central objective at the Large Hadron Collider~(LHC) and future collider facilities. Compared with the Higgs couplings to electroweak~(EW) gauge bosons or third-generation fermions, its self-couplings remain weakly bounded by measurements. Consequently, determining the Higgs trilinear and quartic self-couplings has become a central goal of the long-term physics program at high-energy collider experiments, attracting significant effort from both the experimental and theoretical particle-physics communities.

In the Standard~Model~(SM), the purely bosonic sector is described by the Lagrangian
\beq \label{eq:lag1}
{\cal L} = (D_{\mu}\phi)^{\dagger}(D^{\mu}\phi) + \mu^{2} \hspace{0.25mm} (\phi^{\dagger}\phi) - \lambda \hspace{0.25mm} (\phi^{\dagger}\phi)^{2} \,,
\eeq
where the Higgs field $\phi$ is an $SU(2)_L$ doublet, and $D_{\mu}$ denotes the covariant derivative. The coefficient~$\mu^2$ of the quadratic term is chosen to be positive so that the EW~symmetry is spontaneously broken. After symmetry breaking, the Higgs field can be written in the unitary gauge as
\beq \label{eq:phi}
\phi = \frac{1}{\sqrt{2}} \begin{pmatrix} 0 \\ v + H \end{pmatrix} \,. 
\eeq
Here, $v = \sqrt{\mu^2/\lambda} \simeq 246 \, {\rm GeV}$ is the vacuum expectation value~(VEV) of the Higgs field, and $H$ denotes the physical Higgs boson. In the broken phase the Lagrangian~(\ref{eq:lag1}) takes the form 
\beq \label{eq:lag2}
{\cal L} \supset m_W^2 W_\mu^+ W^{-\mu} + \frac{1}{2} \hspace{0.25mm} m_Z^2 Z_\mu Z^\mu - \frac{1}{2} \hspace{0.25mm} m_H^2 H^2 - v \lambda H^3 - \frac{\lambda}{4} \hspace{0.25mm} H^4 \,, 
\eeq
where the kinetic terms of the Higgs and EW~bosons, as well as the interaction terms between the Higgs and the EW~gauge bosons, have been omitted for simplicity. 

In terms of the fundamental parameters of the SM Lagrangian, the EW~gauge boson and Higgs masses are expressed as 
\beq \label{eq:masses}
 m_W = \frac{g_2 \hspace{0.125mm} v}{2} \,, \qquad m_Z = \frac{\sqrt{g_1^2 + g_2^2} \hspace{0.5mm} v}{2} \,, \qquad m_H = \sqrt{2 \lambda} \hspace{0.25mm} v \,.
\eeq
Here, $g_1$ and $g_2$ denote the gauge couplings of $U(1)_Y$ and $SU(2)_L$, respectively. Once these couplings are fixed by measurements of the $W$- and $Z$-boson interactions with the SM fermions, the measured masses of the $W$, $Z$, and Higgs bosons are sufficient to fully determine the Lagrangian~(\ref{eq:lag2}) and, in turn, all purely bosonic interactions of the SM. In particular, the values of $m_W$ and/or $m_Z$ can be used to determine the Higgs vacuum expectation value $v$, and a measurement of $m_H$ then fixes the quartic coupling $\lambda$. This also fixes the dimensionless Higgs trilinear and quartic self-couplings to their SM values
\beq \label{eq:SM67}
\lambda_3^{\rm SM} = \lambda \,, \qquad \lambda_4^{\rm SM} = \frac{\lambda}{4} \,.
\eeq
Notice also that the Lagrangian~(\ref{eq:lag2}) contains no quintic or higher-order Higgs self-couplings, due to the renormalizability of~the~Lagrangian~(\ref{eq:lag1}), which allows at most a quartic $(\phi^\dagger \phi)^2$ term. In view of this feature as well as Eqs.~(\ref{eq:SM67}), probing the structure of the Higgs potential and the dynamics of EW~symmetry breaking provides a stringent test of the consistency of the~SM. Any observed deviation from the SM prediction could signal the presence of new physics, such as extended scalar sectors, composite Higgs scenarios, or early-universe phenomena like EW~baryogenesis. A broad overview of such beyond-the-SM~(BSM) theories can be found, for example, in~Refs.~\cite{DiMicco:2019ngk,Durieux:2022hbu}.

At the LHC, double-Higgs production provides a direct probe of the Higgs trilinear self-coupling, while loop-induced corrections to single-Higgs production and decay processes~\cite{Gorbahn:2016uoy,Degrassi:2016wml,Bizon:2016wgr,Maltoni:2017ims,Gorbahn:2019lwq,Haisch:2021hvy,Gao:2023bll,Haisch:2024nzv,Ghosh:2025fma} offer complementary indirect constraints. Indeed, both the ATLAS and CMS collaborations have already placed constraints on the Higgs trilinear self-coupling using inclusive measurements of these two classes of processes, based on the complete $\sqrt{s} = 13 \, {\rm TeV}$ dataset corresponding to approximately $140 \, {\rm fb}^{-1}$ of integrated luminosity~\cite{ATLAS:2022jtk,CMS:2024awa}. If interpreted in the $\kappa$-framework~\cite{LHCHiggsCrossSectionWorkingGroup:2012nn,LHCHiggsCrossSectionWorkingGroup:2013rie,LHCHiggsCrossSectionWorkingGroup:2016ypw}, which parameterizes modifications of the generalized Higgs trilinear self-coupling $-v \lambda_3 H^3$ by 
\beq \label{eq:kappa3}
\kappa_3 = \frac{\lambda_3}{\lambda_{3}^{\rm SM}} \,, 
\eeq
with $\lambda_{3}^{\rm SM}$ defined in~Eqs.~(\ref{eq:SM67}), and assuming that all other Higgs couplings take their SM values, the resulting $95\%$ confidence level~(CL) bound on $\kappa_3$ is~\cite{CMS-PAS-HIG-25-014} 
\beq \label{eq:kappa3bounds}
-0.71 < \kappa_3 < 6.1 \,.
\eeq
The high-luminosity LHC~(HL-LHC) is expected to improve the existing bounds significantly, with the latest combined ATLAS and CMS projection based on a dataset of $6 \, {\rm ab}^{-1}$ reporting the $95\%$~CL limit $0.5 < \kappa_3 < 1.7$~\cite{ATL-PHYS-PUB-2025-018}. This hypothetical bound again assumes that all other Higgs couplings take their SM values. Furthermore, since it is derived from theoretical predictions that do not include next-to-leading order~(NLO)~EW~corrections, it is insensitive to the value of the Higgs quartic self-coupling.

Constraining the generalized Higgs quartic self-coupling $-\lambda_4 H^4$, or equivalently
\beq \label{eq:kappa4}
\kappa_4 = \frac{\lambda_4}{\lambda_{4}^{\rm SM}} \,,
\eeq
with $\lambda_{4}^{\rm SM}$ defined in~Eqs.~(\ref{eq:SM67}), is, compared to the already challenging collider tests of $\kappa_3$, much more difficult. While triple-Higgs production provides a direct probe of the Higgs quartic self-coupling, the tiny triple-Higgs production cross section of about $0.11 \, {\rm fb}$ at $\sqrt{s} = 14 \, {\rm TeV}$ --- roughly 300 times smaller than the double-Higgs production cross section of approximately~$36 \, {\rm fb}$ --- makes measuring or constraining $\kappa_4$ extremely challenging. Given these challenges, experimental constraints on $\kappa_4$ did not exist until recently. This has changed with the first searches for triple-Higgs production conducted by the ATLAS~\cite{ATLAS:2024xcs} and CMS~\cite{CMS-PAS-HIG-24-012} collaborations. Under the assumption that all other Higgs couplings take their SM values, in particular~$\kappa_3 = 1$, the resulting bounds on $\kappa_4$ are 
\beq \label{eq:kappa4bound}
-230 < \kappa_4 < 240 \;\;\; (\rm{ATLAS}) \,, \qquad 
-190 < \kappa_4 < 190 \;\;\; (\rm{CMS}) \,,
\eeq
at $95\%$~CL. Under the same assumptions used to obtain~Eq.~(\ref{eq:kappa4bound}), we derive in Section~\ref{sec:SMEFT} a~projected HL-LHC constraint of $-81 < \kappa_4 < 89$, assuming an integrated luminosity of $6~{\rm ab}^{-1}$.

Given the very loose limits on $\kappa_4$, one naturally wonders whether relevant constraints on the Higgs quartic self-coupling can also be obtained from other Higgs production processes once higher-order~EW~corrections are included. An obvious candidate is double-Higgs production, where two-loop~EW~corrections introduce a sensitivity to the Higgs quartic self-coupling. In fact, using effective field theories~(EFTs), three independent groups~\cite{Bizon:2018syu,Borowka:2018pxx,Li:2024iio} have computed such effects.\footnote{Two-loop EW~corrections in the SM have been discussed in~Refs.~\cite{Muhlleitner:2022ijf,Davies:2022ram,Davies:2023npk,Bi:2023bnq,Heinrich:2024dnz,Davies:2024cxd,Bonetti:2025vfd,Bhattacharya:2025egw}.} The first two works~\cite{Bizon:2018syu,Borowka:2018pxx} employ the SM Effective Field Theory~(SMEFT), while the third~\cite{Li:2024iio} uses the Higgs~Effective Field Theory~(HEFT) framework. The main goal of this note is to summarize and compare the key findings of these publications. To this end, we provide a brief description of the SMEFT calculations~\cite{Bizon:2018syu,Borowka:2018pxx} of the $\kappa_3$ and $\kappa_4$ dependence of double-Higgs production in gluon-gluon fusion~(ggF), including NLO~QCD and EW~effects in~Section~\ref{sec:SMEFT}, while Section~\ref{sec:HEFT} offers a more detailed description of the corresponding HEFT computation~\cite{Li:2024iio}, its extension to double-Higgs production in vector-boson fusion~(VBF), and their phenomenological implications. We conclude in~Section~\ref{sec:conclusions}, where we discuss the conceptual differences between the two calculational frameworks and their model dependence. The~complementarity of double- and triple-Higgs production in constraining the Higgs potential is also emphasized. Supplementary material is provided in~Appendix~\ref{app:SMEFT} and~Appendix~\ref{app:SMEFTHEFT}.

\section[Constraining the Higgs potential in the SMEFT]{Constraining the Higgs potential in the SMEFT\protect\footnote{{\it Section authors:} Zach Gillis, Ulrich Haisch, Davide Pagani, Brian Moser, Luca Rottoli, Ambresh Shivaji, and Philipp Windischhofer.} }
\label{sec:SMEFT}

After EW~symmetry breaking, the Higgs field $H$ exhibits non-derivative self-interactions, which can be parametrized in a model-independent form as
\beq \label{eq:V}
V = \frac{1}{2} \hspace{0.25mm} m_H^2 H^2 + \kappa_3 \hspace{0.25mm} \lambda \hspace{0.25mm} v \hspace{0.25mm} H^3 + \kappa_4 \hspace{0.25mm} \frac{\lambda}{4} \hspace{0.25mm} H^4 + \sum_{n=5}^{\infty} \kappa_n \hspace{0.5mm} \frac{\lambda}{v^{n-4}} \hspace{0.5mm} H^n \,,
\eeq
where $\kappa_3$ and $\kappa_4$ are defined as in~Eqs.~(\ref{eq:SM67}), (\ref{eq:kappa3}), and (\ref{eq:kappa4}). The normalization in~Eq.~(\ref{eq:V}) is chosen such that in the SM, one has $\kappa_3 = \kappa_4 = 1$, and $\kappa_n = 0$ for all $n \ge 5$. As explained in the introduction, although constraints on $\kappa_3$ are expected to improve substantially with double-Higgs production data at the HL-LHC, placing direct bounds on the Higgs quartic self-coupling modifier~$\kappa_4$ through measurements of triple-Higgs production remains exceptionally challenging at both future hadron and lepton colliders~\cite{Bizon:2018syu,Borowka:2018pxx,Maltoni:2018ttu,Liu:2018peg,Chiesa:2020awd,Gonzalez-Lopez:2020lpd,Stylianou:2023xit,Papaefstathiou:2023uum,Abouabid:2024gms,Dong:2025lkm}.

Prompted by this observation, Refs.~\cite{Bizon:2018syu,Borowka:2018pxx} evaluated the NLO~EW~corrections induced by the Higgs self-couplings to double-Higgs production through ggF, taking into account the coupling modifiers $\kappa_3$, $\kappa_4$, and $\kappa_5$. Both calculations make use of the relation 
\beq \label{eq:kappa5}
\kappa_5 = \frac{7}{4} - \frac{9}{4} \hspace{0.5mm} \kappa_3 + \frac{1}{2} \hspace{0.5mm} \kappa_4 \,, 
\eeq 
which is derived under the assumption that, within the SMEFT framework 
\beq \label{eq:LSMEFT}
{\cal L}_{\rm SMEFT} \supset \frac{C_6}{\Lambda^2} \, Q_6 + \frac{C_8}{\Lambda^4} \, Q_8 \,, 
\eeq 
only the dimension-six and dimension-eight operators 
\beq \label{eq:dim67}
Q_6 = (\phi^\dagger \phi)^3 \,, \qquad Q_8 = (\phi^\dagger \phi)^4 \,,
\eeq
contribute to modifications of the Higgs self-couplings, while other possible operators, such as the dimension-ten operator $Q_{10} = (\phi^\dagger \phi)^5$, are assumed to have no effect. In Eq.~(\ref{eq:LSMEFT}), the symbol $\Lambda$ denotes the scale of new physics that suppresses the higher-dimensional operators~$Q_6$ and~$Q_8$, rendering their Wilson coefficients $C_6$ and $C_8$ dimensionless. In Appendix~\ref{app:SMEFT}, we show that restricting the SMEFT Lagrangian to operators of the form $(\phi^\dagger \phi)^n$, as in Eq.~(\ref{eq:LSMEFT}), provides a good approximation for BSM models that primarily modify the Higgs self-couplings, while inducing only minor effects on other SMEFT operators. Strictly speaking, the phenomenological analysis below therefore applies only to such models. These scenarios are, however, particularly interesting, as probing the Higgs self-couplings via multi-Higgs production at the HL-LHC and future colliders explores largely uncharted parameter space, offering genuine discovery potential even if single-Higgs observables appear SM-like.
 
It is important to emphasize that, under the above assumption, a one-to-one correspondence exists between the coupling modifiers $\kappa_3$ and $\kappa_4$ and the Wilson coefficients $C_6$ and $C_8$ introduced in~Eq.~(\ref{eq:LSMEFT}). This mapping is explicitly given by
\beq \label{eq:C67}
\begin{split}
C_6 & = -\frac{2 \hspace{0.125mm} m_h^2 \hspace{0.25mm} \Lambda^2}{v^4} \, \big ( \kappa_3 - 1 \big ) + \frac{m_h^2 \hspace{0.25mm} \Lambda^2}{4 \hspace{0.125mm} v^4} \, \big ( \kappa_4 - 1 \big ) \,, \\[2mm]
C_8 & = \frac{3 \hspace{0.125mm} m_h^2 \hspace{0.25mm} \Lambda^4}{4 \hspace{0.125mm} v^6} \, \big ( \kappa_3 - 1 \big ) - \frac{m_h^2 \hspace{0.25mm} \Lambda^4}{8 \hspace{0.125mm} v^6} \, \big ( \kappa_4 - 1 \big ) \,.
\end{split}
\eeq
In the following, all formulas, constraints, and limits in this section are expressed in terms of $\kappa_3$ and $\kappa_4$. Alternatively, one may work with $C_6$ and $C_8$, using the correspondence given in~Eqs.~(\ref{eq:C67}).

Employing~Eqs.~(\ref{eq:kappa5}) and (\ref{eq:LSMEFT}), Ref.~\cite{Bizon:2018syu} computed the NLO~EW~corrections to the $gg \to HH$ amplitudes at ${\cal O} (\kappa_4)$ and ${\cal O} (\kappa_3 \kappa_4)$, while the computation in~Ref.~\cite{Borowka:2018pxx} additionally includes contributions at ${\cal O} (\kappa_3^2)$ and~${\cal O} (\kappa_3^3)$. Subsequently, NLO~EW~corrections involving $\kappa_3$ and $\kappa_4$ have been computed within the HEFT framework~\cite{Li:2024iio}. Details of this calculation are provided in~Section~\ref{sec:HEFT}. The corrections computed in~Ref.~\cite{Bizon:2018syu} have been incorporated into the \texttt{POWHEG BOX} framework~\cite{Alioli:2010xd}, where they are combined with NLO~QCD corrections that account for $\kappa_3$ variations, including the full top-quark mass effects~\cite{Borowka:2016ypz,Borowka:2016ehy,Heinrich:2017kxx,Heinrich:2019bkc,Bagnaschi:2023rbx}. As shown in~Ref.~\cite{Bizon:2024juq}, the \texttt{POWHEG BOX} implementation accurately reproduces the LHC Higgs Cross Section Working Group’s current recommendations for the double-Higgs production cross section as a function of $\kappa_3$, based on~Refs.~\cite{Grazzini:2018bsd,Amoroso:2020lgh}, which carry relative uncertainties of about $\pm 15\%$ arising from the choice of the renormalization and top-quark mass schemes~\cite{Baglio:2020wgt}.

The \texttt{POWHEG BOX} implementation described above enables an efficient computation of the inclusive $gg \to HH$ production cross section at the LHC. Using the \texttt{PDF4LHC15\_NLO} parton distribution functions~(PDFs)~\cite{Butterworth:2015oua},\footnote{The specific choice of PDFs has only a very minor numerical impact on all the formulas and results presented in this and the following section.} with $m_t = 173 \, {\rm GeV}$ and adopting the scale choice $\mu_R = \mu_F = m_{HH}/2$ — where $m_t$ denotes the top-quark mass, $\mu_R$ and $\mu_F$ represent the renormalization and factorization scales, respectively, and $m_{HH}$ is the invariant mass of the final-state Higgs pair — the following signal strengths for collisions at $\sqrt{s} = 13 \, {\rm TeV}$, $\sqrt{s} = 13.6 \, {\rm TeV}$, and $\sqrt{s} = 14 \, {\rm TeV}$ are obtained~\cite{Bizon:2024juq}:
\beq \label{eq:muhhs}
\begin{split}
\mu_{2h, \mathrm{SMEFT}}^{13 \, {\rm TeV}} & = 2.21 - 1.54 \hspace{0.25mm} \kappa_3 - 1.04 \cdot 10^{-3} \hspace{0.25mm} \kappa_4 + 3.35 \cdot 10^{-1} \hspace{0.25mm} \kappa_3^2 + 4.06 \cdot 10^{-3} \hspace{0.25mm} \kappa_3 \hspace{0.25mm} \kappa_4 \\[1mm]
& \phantom{xx} + 5.59 \cdot 10^{-5} \hspace{0.25mm} \kappa_4^2 - 1.62 \cdot 10^{-3} \hspace{0.25mm} \kappa_3^2 \hspace{0.25mm} \kappa_4 - 3.86 \cdot 10^{-5} \hspace{0.25mm} \kappa_3 \hspace{0.25mm} \kappa_4^2 + 9.82 \cdot 10^{-6} \hspace{0.25mm} \kappa_3^2 \hspace{0.25mm} \kappa_4^2 \,, \\[2mm]
\mu_{2h, \mathrm{SMEFT}}^{13.6 \, {\rm TeV}} & = 2.20 - 1.53 \hspace{0.25mm} \kappa_3 - 1.01 \cdot 10^{-3} \hspace{0.25mm} \kappa_4 + 3.32 \cdot 10^{-1} \hspace{0.25mm} \kappa_3^2 + 4.02 \cdot 10^{-3} \hspace{0.25mm} \kappa_3 \hspace{0.25mm} \kappa_4 \\[1mm]
& \phantom{xx} + 5.60 \cdot 10^{-5} \hspace{0.25mm} \kappa_4^2 - 1.60 \cdot 10^{-3} \hspace{0.25mm} \kappa_3^2 \hspace{0.25mm} \kappa_4 - 3.85 \cdot 10^{-5} \hspace{0.25mm} \kappa_3 \hspace{0.25mm} \kappa_4^2 + 9.77 \cdot 10^{-6} \hspace{0.25mm} \kappa_3^2 \hspace{0.25mm} \kappa_4^2 \,, \\[2mm]
\mu_{2h, \mathrm{SMEFT}}^{14 \, {\rm TeV}} & = 2.20 - 1.53 \hspace{0.25mm} \kappa_3 - 0.98 \cdot 10^{-3} \hspace{0.25mm} \kappa_4 + 3.30 \cdot 10^{-1} \hspace{0.25mm} \kappa_3^2 + 4.00 \cdot 10^{-3} \hspace{0.25mm} \kappa_3 \hspace{0.25mm} \kappa_4 \\[1mm]
& \phantom{xx} + 5.60 \cdot 10^{-5} \hspace{0.25mm} \kappa_4^2 - 1.59 \cdot 10^{-3} \hspace{0.25mm} \kappa_3^2 \hspace{0.25mm} \kappa_4 - 3.84 \cdot 10^{-5} \hspace{0.25mm} \kappa_3 \hspace{0.25mm} \kappa_4^2 + 9.74 \cdot 10^{-6} \hspace{0.25mm} \kappa_3^2 \hspace{0.25mm} \kappa_4^2 \,. 
\end{split}
\eeq
Three features of the above expressions are worth highlighting. First, each power of $\kappa_4$ introduces a relative suppression of ${\cal O} (10^{-3})$, reflecting the fact that $\kappa_4$ contributions enter only at NLO, whereas the leading $\kappa_3$ dependence appears already at leading order~(LO) in EW~interactions. Second, the different coefficients show only a very weak dependence on $\sqrt{s}$ across the range of center-of-mass~(CM) energies considered. Additional expressions analogous to~Eqs.~(\ref{eq:muhhs}), corresponding to different scale choices at the LHC and relevant to the hadron mode of the Future Circular Collider~(FCC-hh), can be found in~Refs.~\cite{Bizon:2024juq,Haisch:2025pql}. Using the results of Ref.~\cite{Haisch:2025pql}, one can also include next-to-next-to-leading order~(NNLO)~EW~corrections in~Eqs.~(\ref{eq:muhhs}), arising from two-loop contributions to the wave function renormalization constant of the Higgs field. These corrections generally have a limited numerical impact. For~example, in the case of the $\kappa_4^2$ coefficients, NNLO~EW~effects due to the Higgs wave function renormalization correspond to a relative shift of about $20\%$. The same NNLO~EW effects from Higgs wave function renormalization also induce a dependence of all single-Higgs production processes on $\kappa_4$. However, since these effects first arise at three loops, the resulting sensitivity to $\kappa_4$ is very limited~\cite{Haisch:2025pql}.

\begin{figure}[t!]
\begin{center}
\includegraphics[height=0.45\textwidth]{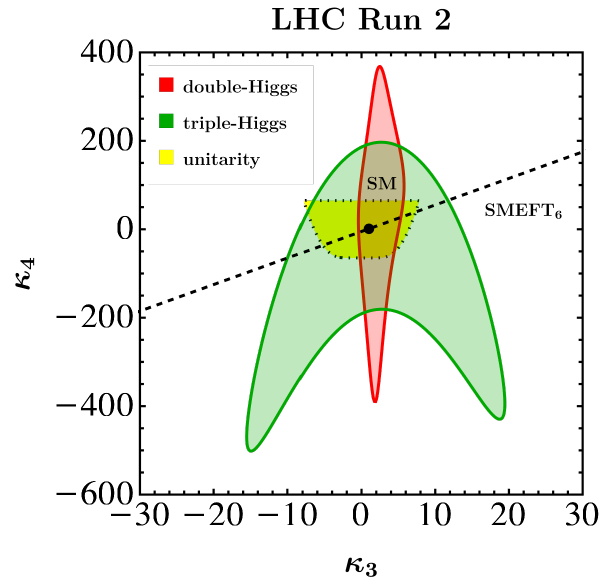} \qquad 
\includegraphics[height=0.45\textwidth]{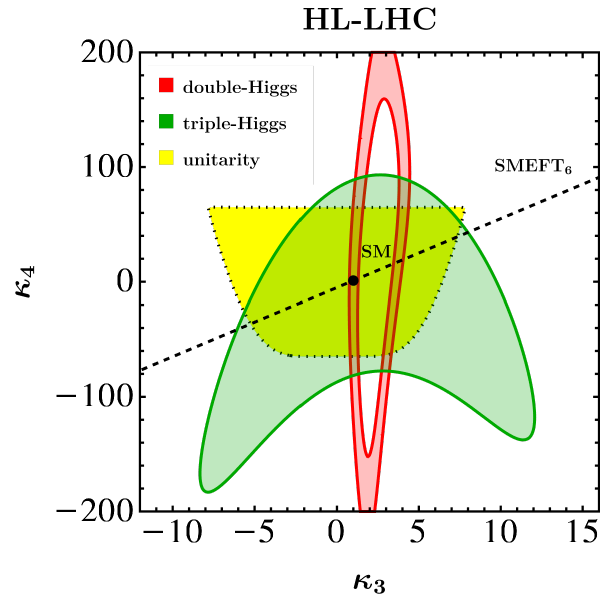}
\end{center}
\vspace{-4mm} 
\caption{\label{fig:planes} Constraints in the $\kappa_3\hspace{0.25mm}$--$\hspace{0.25mm}\kappa_4$~plane at the~LHC~Run~2~(left) and the~HL-LHC~(right). The red and green contours correspond to the preferred $68\%$~CL regions that arise from inclusive double- and triple-Higgs production, respectively. The SM is indicated by the black points, and the black dashed lines correspond to $\kappa_4 - 1 = 6 \, \big ( \kappa_3 - 1 \big )$, i.e., the relation between $\kappa_3$ and $\kappa_4$ that holds in the SMEFT at the level of dimension-six operators. The yellow regions, outlined with black dotted lines, indicate the constraint from perturbative unitarity derived from tree-level $HH \to HH$ scattering.}
\end{figure}

The two panels in~Figure~\ref{fig:planes} present the results of a sensitivity analysis for constraining $\kappa_3$ and~$\kappa_4$, based on both existing LHC~Run~2 data and projections for the HL-LHC with an assumed integrated luminosity of $3 \, {\rm ab}^{-1}$. The red and green contours indicate the $68\%$~CL limits derived from double- and triple-Higgs production, respectively. Constraints from double-Higgs production are obtained using Eqs.~(\ref{eq:muhhs}), while those from triple-Higgs production are computed at LO in QCD using \texttt{MadGraph5\_aMC@NLO}~\cite{Alwall:2014hca}, with NLO~QCD corrections from~Ref.~\cite{Maltoni:2014eza} included as an overall normalization factor. Expressions for signal strengths in triple-Higgs production, analogous to Eqs.~(\ref{eq:muhhs}), can be found in~Refs.~\cite{Bizon:2018syu,Haisch:2025pql}. The double- and triple-Higgs constraints shown for LHC~Run~2 assume $\mu_{2h}^{\textrm{LHC~Run~2}} < 2.9$~\cite{ATLAS:2024ish} and $\mu_{3h}^{\textrm{LHC~Run~2}} < 588$~\cite{CMS-PAS-HIG-24-012}, while at the HL-LHC we have assumed that the corresponding signal strengths can be constrained to $0.77 < \mu_{2h}^{\textrm{HL-LHC}} < 1.23$ and $\mu_{3h}^{\textrm{HL-LHC}} < 125$. These hypothetical limits are consistent with those derived in Ref.~\cite{ATL-PHYS-PUB-2022-005} and Ref.~\cite{ATL-PHYS-PUB-2025-003}, respectively. The SM prediction is marked by the black dots, while the dashed black lines represent the set of solutions satisfying $\kappa_4 - 1 = 6 \, \big ( \kappa_3 - 1 \big )$. This relation holds under the assumption that $Q_6$ in~Eqs.~(\ref{eq:LSMEFT}) is the only SMEFT operator with a non-zero Wilson coefficient.

The results shown in the figure clearly demonstrate that, at the LHC, the constraints from double- and triple-Higgs production are largely complementary. Double-Higgs production is primarily sensitive to variations in $\kappa_3$, whereas triple-Higgs production generally yields stronger bounds on $\kappa_4$. This~observation has also been made in~Refs.~\cite{Bizon:2018syu,Borowka:2018pxx,Li:2024iio,Bizon:2024juq,Haisch:2025pql}. Assuming $\kappa_3 = 1$, our analysis yields the $95\%$~CL constraint 
\beq \label{eq:LHCRun2k4}
-185 < \kappa_4 < 193 \,,
\eeq
consistent with the limits reported by the ATLAS and CMS collaborations in~Refs.~\cite{ATLAS:2024xcs,CMS-PAS-HIG-24-012} and quoted in~Eqs.~(\ref{eq:kappa4bound}). Our~HL-LHC projection, assuming an integrated luminosity of $6\, {\rm ab}^{-1}$, reduces the allowed $95\%$ CL range to
\beq \label{eq:HLLHCk4}
-81 < \kappa_4 < 89 \,,
\eeq 
comparable to the limits set by perturbative unitarity~\cite{Liu:2018peg,Stylianou:2023xit,DiLuzio:2017tfn}, depicted as a yellow region with a black dotted outline in both panels of Figure~\ref{fig:planes}. Improvements in background uncertainties and $b$-tagging performance relative to the LHC~Run~2 baseline could further strengthen the bound given in~Eq.~(\ref{eq:HLLHCk4}). We have confirmed that, within this region, the two NLO~EW~computations of double-Higgs production~\cite{Bizon:2018syu,Borowka:2018pxx} yield numerically similar constraints in the $\kappa_3\hspace{0.25mm}$--$\hspace{0.25mm}\kappa_4$ plane.\footnote{Since the two computations include different subsets of NLO~EW~corrections induced by the Higgs self-couplings, a detailed comparison is non-trivial and is therefore deferred to future work.} Based on perturbativity considerations~\cite{Maltoni:2018ttu,DiLuzio:2017tfn}, the calculation in~Ref.~\cite{Borowka:2018pxx} is valid only within the range $-4 \lesssim \kappa_3 \lesssim 6$ and $6 \kappa_3 - 36 \lesssim \kappa_4 \lesssim 6 \kappa_3 + 26$. Although this region is not shown in~Figure~\ref{fig:planes}, it encompasses the overlap of the red and yellow areas corresponding to constraints from double-Higgs production and unitarity, respectively. Importantly, if only the value of the total cross section of double-Higgs production is taken into account, two distinct regions in parameter space are likely to remain allowed, with one centered around the SM point and the other near $\{\kappa_3, \kappa_4\} \simeq \{3.5, 16\}$. Note that the LHC Run~2 analysis exhibits the same BSM solutions, but they are not resolved because the double-Higgs signal strength is only bounded from above, $\mu_{2}^{\textrm{LHC~Run~2}} < 2.9$. At the HL-LHC, with $0.77 < \mu_{2h}^{\textrm{HL-LHC}} < 1.23$, and since~Eqs.~(\ref{eq:muhhs}) are polynomial in $\kappa_3$ and $\kappa_4$, this produces an elongated ellipsoidal band rather than a closed contour. Furthermore, observe that this BSM solution is consistent with the $\kappa_3$ bound quoted after Eqs.~(\ref{eq:kappa3bounds}) from ATLAS and CMS projections. In addition to the total cross section, these projections incorporate differential information via Monte Carlo~(MC) simulations, enabling them to break the degeneracy and favor values near the SM point. We also notice that even at the HL-LHC the total cross section alone will not suffice to distinguish between BSM scenarios in which deviations in Higgs self-interactions arise exclusively from the SMEFT operator $Q_6$, or from a combination of the operators $Q_6$ and $Q_8$ as defined in~Eqs.~(\ref{eq:dim67}). On the other hand, incorporating kinematic information in double-Higgs production could affect this conclusion --- see~Section~\ref{sec:HEFT}. As~shown in Refs.~\cite{Bizon:2018syu,Haisch:2025pql}, the FCC-hh would be able to resolve the~degeneracy depicted in the right plot in Figure~\ref{fig:planes}.

\section[Constraining the Higgs potential in the HEFT]{Constraining the Higgs potential in the HEFT\protect\footnote{{\it Section authors:} Jia-Le Ding, Hai Tao Li, Zong-Guo Si, Jian Wang, Xiao Zhang, and Dan Zhao.} }
\label{sec:HEFT}

Before discussing in some detail the calculation of the NLO~EW~corrections to double-Higgs production in the HEFT, it is useful to comment on the impact of QCD corrections on the extraction of~$\kappa_3$ from double-Higgs production measurements. Keeping the~$\kappa_{3}$ dependence, the LO~cross section for double-Higgs production via~ggF can be written as 
\beq \label{eq:ggFLO} 
\sigma_{\rm ggF, LO}^{\sqrt{s}} = \big ( \sigma_{\rm ggF, LO}^{\sqrt{s}} \big )_{\rm SM}
\left( 1 - 0.81 \, \big ( \kappa_3 - 1 \big ) + 0.28 \, \big ( \kappa_3 - 1 \big )^{2} \right) \,,
\eeq
where
\beq \label{eq:LOSMggF} 
\big ( \sigma_{\rm ggF, LO}^{13 \, {\rm TeV}} \big )_{\rm SM} = 16.7 \, {\rm fb} \,, \qquad
\big ( \sigma_{\rm ggF, LO}^{13.6 \, {\rm TeV}} \big)_{\rm SM} = 18.6 \, {\rm fb} \,, \qquad
\big ( \sigma_{\rm ggF, LO}^{14 \, {\rm TeV}} \big )_{\rm SM} = 19.8 \, {\rm fb} \,,
\eeq
are the LO SM cross-section predictions at the corresponding CM energies. Higher-order QCD corrections do not alter the polynomial structure of these expressions, but only modify the coefficients. For the corresponding NLO~QCD cross sections, we find
\beq \label{eq:ggFNLO} 
\sigma_{\rm ggF, NLO}^{\sqrt{s}} = \big ( \sigma_{\rm ggF, NLO}^{\sqrt{s}} \big )_{\rm SM}
\left( 1 - 0.87 \, \big ( \kappa_3 - 1 \big ) + 0.33 \, \big ( \kappa_3 - 1 \big )^{2} \right) \,,
\eeq
where
\beq \label{eq:NLOSMggF} 
\big ( \sigma_{\rm ggF, NLO}^{13 \, {\rm TeV}} \big )_{\rm SM} = 27.8 \, {\rm fb} \,, \qquad
\big ( \sigma_{\rm ggF, NLO}^{13.6 \, {\rm TeV}} \big )_{\rm SM} = 30.8 \, {\rm fb} \,, \qquad
\big ( \sigma_{\rm ggF, NLO}^{14 \, {\rm TeV}} \big )_{\rm SM} = 33.0 \, {\rm fb} \,.
\eeq
These NLO~QCD predictions were obtained using the two-loop $gg \to HH$ amplitude~\cite{Borowka:2016ypz} as implemented in the~\texttt{POWHEG~BOX}, while the one-loop amplitudes for $gg \to HH g$ and other related partonic processes were generated with \texttt{OpenLoops}~\cite{Buccioni:2019sur,Buccioni:2017yxi}. Infrared divergences are treated using the dipole subtraction method~\cite{Catani:1996vz,Gleisberg:2007md}. The NLO~QCD results have also been cross-checked by means of the~\texttt{ggxy}~program~\cite{Davies:2023vmj,Davies:2025qjr}. The following comment concerning Eqs.~(\ref{eq:ggFLO}) and~(\ref{eq:ggFNLO}) is in order. We find that the $\kappa_{3}$ dependence is essentially independent of $\sqrt{s}$ at each perturbative order in QCD. However, the $\kappa_{3}$ dependence does change when going from LO to NLO in QCD, with the linear and quadratic terms in $\kappa_3$ being modified differently. While the linear term receives a relative correction of about $+7\%$, the quadratic term is shifted by approximately $+18\%$. Including QCD corrections therefore enhances the sensitivity to modifications of the Higgs trilinear self-coupling, making it important to account for them when interpreting LHC data on ggF double-Higgs production.

Compared to the ggF channel, the impact of QCD corrections is much less pronounced in VBF double-Higgs production. We obtain
\beq \label{eq:VBF}
\begin{split}
\sigma_{{\rm VBF}, \rm LO}^{13 \, {\rm TeV}} & = 1.71 
\left( 1 - 0.91 \, \big ( \kappa_3 - 1 \big ) + 0.73 \, \big ( \kappa_3 - 1 \big )^{2} \right) {\rm fb} \,, \\[2mm]
\sigma_{{\rm VBF}, \rm LO}^{13.6 \, {\rm TeV}} & = 1.90
\left( 1 - 0.89 \, \big ( \kappa_3 - 1 \big ) + 0.72 \, \big ( \kappa_3 - 1 \big )^{2} \right) {\rm fb} \,, \\[2mm]
\sigma_{{\rm VBF}, \rm LO}^{14 \, {\rm TeV}} & = 2.03 
\left( 1 - 0.90 \, \big ( \kappa_3 - 1 \big ) + 0.71 \, \big ( \kappa_3 - 1 \big )^{2} \right) {\rm fb} \,, \\[2mm]
\sigma_{{\rm VBF}, \rm N^3LO}^{13 \, {\rm TeV}} & = 1.70 
\left( 1 - 0.89 \, \big ( \kappa_3 - 1 \big ) + 0.72 \, \big ( \kappa_3 - 1 \big )^{2} \right) {\rm fb} \,, \\[2mm]
\sigma_{{\rm VBF}, \rm N^3LO}^{13.6 \, {\rm TeV}} & = 1.89 
\left( 1 - 0.88 \, \big ( \kappa_3 - 1 \big ) + 0.71 \, \big ( \kappa_3 - 1 \big )^{2} \right) {\rm fb} \,, \\[2mm]
\sigma_{{\rm VBF}, \rm N^3LO}^{14 \, {\rm TeV}} & = 2.01 
\left( 1 - 0.88 \, \big ( \kappa_3 - 1 \big ) + 0.71 \, \big ( \kappa_3 - 1 \big )^{2} \right) {\rm fb} \,.
\end{split}
\eeq
One observes that the inclusion of next-to-next-to-next-to-leading order~(N$^3$LO) QCD corrections reduces the SM cross sections by about~$-1\%$ at each CM energy, leaving the $\kappa_3$ dependence essentially unchanged. This is expected from the factorization properties of VBF-like processes, which can be accurately treated as two independent deep-inelastic scattering processes using the structure function approach. The effect of non-factorizable NNLO QCD corrections associated with $\kappa_3$ in VBF double-Higgs production was studied in~Ref.~\cite{Jager:2025isz} and found to be at most approximately $-0.35\%$ for $-2 < \kappa_3 < 4$.

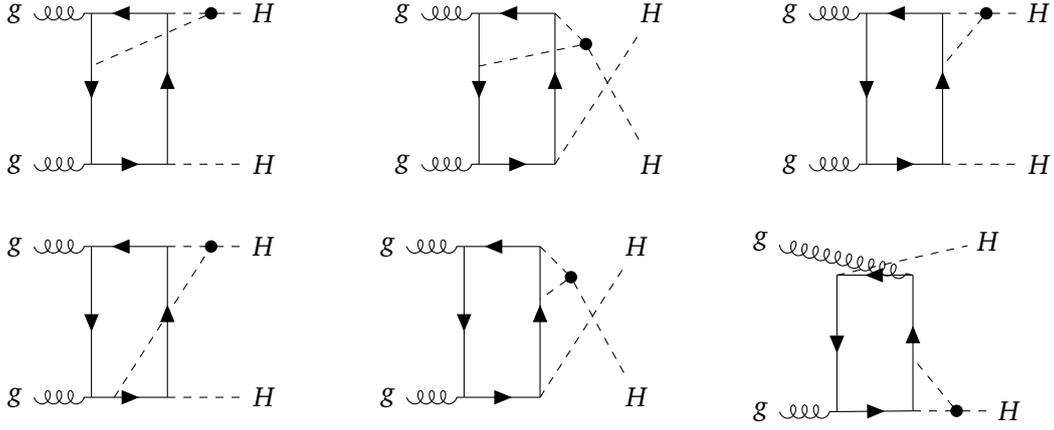
\begin{figure}[t!]
	\begin{minipage}{0.32\linewidth}
		\centering
		\begin{tikzpicture}
		\begin{feynman}
		\vertex (i1) {\(g\)};
		\vertex[right=1 cm of i1] (a);
		\vertex[right=1 cm of a] (b);
		\vertex[right=0.5 cm of b, dot] (e) {};
		\vertex[right=0.7 cm of e] (f1) {\(H\)};
		
		\vertex[below=0.7 cm of a] (f);
		\vertex[below=1.3 cm of f] (c);
		\vertex[below=2 cm of i1] (i2) {\(g\)};
		\vertex[right=1 cm of c] (d);
		\vertex[right=1 cm of d] (f2) {\(H\)};
		
		\diagram* {
			(i1) -- [gluon] (a) -- [anti fermion] (b) -- [scalar] (e)[dot] -- [scalar] (f1),
			(i2) -- [gluon] (c) -- [fermion] (d) -- [scalar] (f2),
			(a) -- [fermion] (c),
			(b) -- [anti fermion] (d),
			(e) -- [scalar] (f)};
		\node at (2.1,-2.6) {};
		\end{feynman}
		\end{tikzpicture}
	\end{minipage}
	\begin{minipage}{0.32\linewidth}	
		\centering
		\begin{tikzpicture}
		\begin{feynman}
		\vertex (i1) {\(g\)};
		\vertex[right=1 cm of i1] (a);
		\vertex[right=1 cm of a] (b);
		\vertex[below right=0.5 cm of b, dot] (e) {};
		\vertex[right=1 cm of b] (f1) {\(H\)};
		
		\vertex[below=0.7 cm of a] (f);
		\vertex[below=1.3 cm of f] (c);
		\vertex[below=2 cm of i1] (i2) {\(g\)};
		\vertex[right=1 cm of c] (d);
		\vertex[right=1 cm of d] (f2) {\(H\)};
		
		\diagram* {
			(i1) -- [gluon] (a) -- [anti fermion] (b) -- [scalar] (e)[dot] -- [scalar] (f2),
			(i2) -- [gluon] (c) -- [fermion] (d) -- [scalar] (f1),
			(a) -- [fermion] (c),
			(b) -- [anti fermion] (d),
			(e) -- [scalar] (f)};
		\node at (1.9,-2.6) {};
		\end{feynman}
		\end{tikzpicture}
	\end{minipage}
	\begin{minipage}{0.32\linewidth}	
		\centering
		\begin{tikzpicture}
		\begin{feynman}
		\vertex (i1) {\(g\)};
		\vertex[right=1 cm of i1] (a);
		\vertex[right=1 cm of a] (b);
		\vertex[right=0.5 cm of b, dot] (e) {};
		\vertex[right=0.7 cm of e] (f1) {\(H\)};
		
		\vertex[below=0.7 cm of b] (f);
		\vertex[below=1.3 cm of f] (d);
		\vertex[below=2 cm of i1] (i2) {\(g\)};
		\vertex[right=1 cm of i2] (c);
		\vertex[right=1 cm of d] (f2) {\(H\)};
		
		\diagram* {
			(i1) -- [gluon] (a) -- [anti fermion] (b) -- [scalar] (e)[dot] -- [scalar] (f1),
			(i2) -- [gluon] (c) -- [fermion] (d) -- [scalar] (f2),
			(a) -- [fermion] (c),
			(b) -- [anti fermion] (d),
			(e) -- [scalar] (f)};
		\node at (1.8,-2.6) {};
		\end{feynman}
		\end{tikzpicture}
	\end{minipage}
	\begin{minipage}{0.32\linewidth}	
		\centering
		\begin{tikzpicture}
		\begin{feynman}
		\vertex (i1) {\(g\)};
		\vertex[right=1 cm of i1] (a);
		\vertex[right=1 cm of a] (b);
		\vertex[right=0.5 cm of b, dot] (e) {};
		\vertex[right=0.7 cm of e] (f1) {\(H\)};
		
		\vertex[below=2 cm of i1] (i2) {\(g\)};
		\vertex[right=1 cm of i2] (c);
		\vertex[right=0.3 cm of c] (d);
		\vertex[right=0.7 cm of d] (f);
		\vertex[right=1 cm of f] (f2) {\(H\)};
		
		\diagram* {
			(i1) -- [gluon] (a) -- [anti fermion] (b) -- [scalar] (f1),
			(i2) -- [gluon] (c) -- [fermion] (f) -- [scalar] (f2),
			(a) -- [fermion] (c),
			(b) -- [anti fermion] (f),
			(d) -- [scalar] (e)[dot]};
		\node at (2.1,-2.6) {};
		\end{feynman}
		\end{tikzpicture}
	\end{minipage}
	\begin{minipage}{0.32\linewidth}	
		\centering
		\begin{tikzpicture}
		\begin{feynman}
		\vertex (i1) {\(g\)};
		\vertex[right=1 cm of i1] (a);
		\vertex[right=1 cm of a] (b);
		\vertex[below right=0.5 cm of b, dot] (e) {};
		\vertex[right=1 cm of b] (f1) {\(H\)};
		
		\vertex[below=0.7 cm of b] (f);
		\vertex[below=1.3 cm of f] (d);
		\vertex[below=2 cm of i1] (i2) {\(g\)};
		\vertex[right=1 cm of i2] (c);
		\vertex[right=1 cm of d] (f2) {\(H\)};
		
		\diagram* {
			(i1) -- [gluon] (a) -- [anti fermion] (b) -- [scalar] (e)[dot] -- [scalar] (f2),
			(i2) -- [gluon] (c) -- [fermion] (d) -- [scalar] (f1),
			(a) -- [fermion] (c),
			(b) -- [anti fermion] (d),
			(e) -- [scalar] (f)};
		\node at (2.1,-2.6) {};
		\end{feynman}
		\end{tikzpicture}
	\end{minipage}
	\begin{minipage}{0.32\linewidth}	
		\centering
		\begin{tikzpicture}
		\begin{feynman}
		\vertex (i1);
		\vertex[above=0.2 cm of i1] (i12) {\(g\)};
		\vertex[right=1 cm of i1] (a);
		\vertex[right=1 cm of a] (b);
		\vertex[right=1 cm of b] (f1);
		\vertex[above=0.2 cm of f1] (f12) {\(H\)};
		
		\vertex[below=1.1 cm of b] (f);
		\vertex[below=0.7 cm of f] (d);
		\vertex[below=1.55 cm of i1] (i2) {\(g\)};
		\vertex[right=1 cm of i2] (c);
		\vertex[right=0.5 cm of d, dot] (e) {};
		\vertex[right=0.7 cm of e] (f2) {\(H\)};
		
		\diagram* {
			(i12) -- [gluon] (b) -- [fermion] (a) -- [scalar] (f12),
			(i2) -- [gluon] (c) -- [fermion] (d) -- [scalar] (e)[dot] -- [scalar] (f2),
			(a) -- [fermion] (c),
			(b) -- [anti fermion] (d),
			(e) -- [scalar] (f)};
		\node at (2.1,-2.3) {};
		\end{feynman}
		\end{tikzpicture}
	\end{minipage}
	\caption{Feynman diagrams contributing to $gg \to HH$ that contain a single Higgs trilinear self-coupling (black dot) and three top-quark Yukawa couplings. The diagrams with reversed fermion flow are omitted.}
	\label{figNLO_HY}
\end{figure}

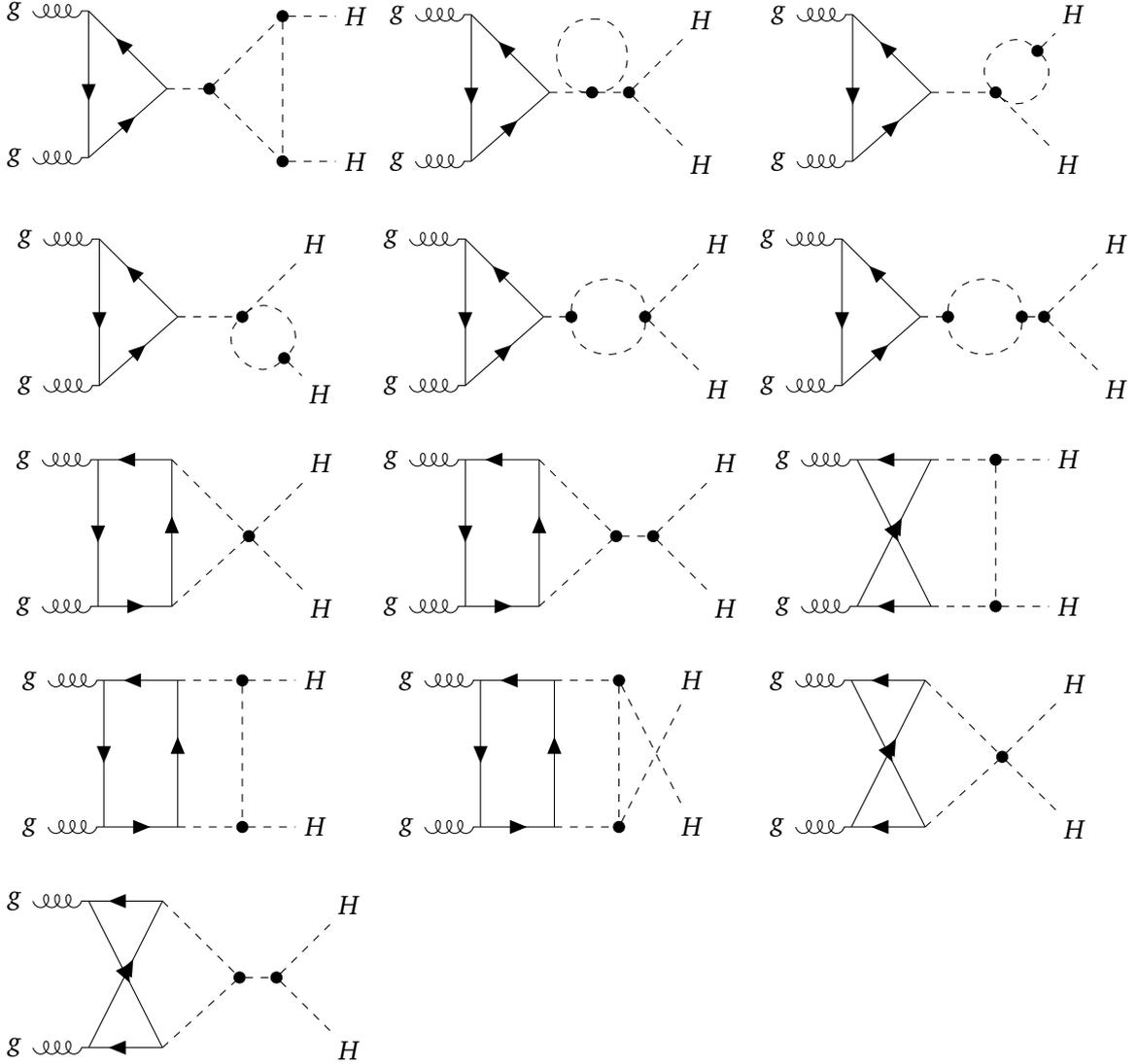
\begin{figure}[t!]
\begin{minipage}{0.32\linewidth}
	\centering
	\begin{tikzpicture}
	\begin{feynman}
	\vertex (i1) {\(g\)};
	\vertex[right=1 cm of i1] (a);
	\vertex[below right=1.5 cm of a] (b);
	\vertex[right=0.5 cm of b, dot] (d) {};
	\vertex[above right=1.4 cm of d, dot] (e) {};
	\vertex[below right=1.4 cm of d, dot] (f) {};
	\vertex[right=1 cm of e] (f1) {\(H\)};
	\vertex[right=1 cm of f] (f2) {\(H\)};
	
	\vertex[below=2 cm of i1] (i2) {\(g\)};
	\vertex[right=1 cm of i2] (c);
	
	\diagram* {
		(i1) -- [gluon] (a) -- [anti fermion] (b) -- [scalar] (d)[dot] -- [scalar] (e)[dot] -- [scalar] (f1),
		(i2) -- [gluon] (c) -- [fermion] (b),
		(c) -- [anti fermion] (a),
		(d) -- [scalar] (f)[dot] -- [scalar] (f2),
		(e) -- [scalar] (f)};
	\node at (2.2,-2.7) {};
	\end{feynman}
	\end{tikzpicture}
\end{minipage}
\begin{minipage}{0.32\linewidth}
	\centering
	\begin{tikzpicture}
	\begin{feynman}
	\vertex (i1) {\(g\)};
	\vertex[right=1 cm of i1] (a);
	\vertex[below right=1.5 cm of a] (b);
	\vertex[right=0.5 cm of b, dot] (d) {};
	\vertex[right=0.5 cm of d, dot] (e) {};
	\vertex[above right=1.4 cm of e] (f1) {\(H\)};
	\vertex[below right=1.4 cm of e] (f2) {\(H\)};
	\vertex[above =1 cm of d] (f);
	
	\vertex[below=2 cm of i1] (i2) {\(g\)};
	\vertex[right=1 cm of i2] (c);
	
	\diagram* {
		(i1) -- [gluon] (a) -- [anti fermion] (b) -- [scalar] (d)[dot] -- [scalar] (e)[dot] -- [scalar] (f1),
		(i2) -- [gluon] (c) -- [fermion] (b),
		(c) -- [anti fermion] (a),
		(d) -- [scalar, half left] (f) -- [scalar, half left] (d),
		(e) -- [scalar] (f2)};
	\node at (2.2,-2.6) {};
	\end{feynman}
	\end{tikzpicture}
\end{minipage}
\begin{minipage}{0.32\linewidth}
	\centering
	\begin{tikzpicture}
	\begin{feynman}
	\vertex (i1) {\(g\)};
	\vertex[right=1 cm of i1] (a);
	\vertex[below right=1.5 cm of a] (b);
	\vertex[right=0.8 cm of b, dot] (d) {};
	\vertex[above right=0.8 cm of d, dot] (e) {};
	\vertex[above right=0.7 cm of e] (f1) {\(H\)};
	\vertex[below right=1.4 cm of d] (f2) {\(H\)};
	
	\vertex[below=2 cm of i1] (i2) {\(g\)};
	\vertex[right=1 cm of i2] (c);
	
	\diagram* {
		(i1) -- [gluon] (a) -- [anti fermion] (b) -- [scalar] (d)[dot] -- [scalar, half left] (e)[dot] -- [scalar, half left] (d),
		(i2) -- [gluon] (c) -- [fermion] (b),
		(c) -- [anti fermion] (a),
		(d) -- [scalar] (f2),
		(e) -- [scalar] (f1)};
	\node at (2.2,-2.6) {};
	\end{feynman}
	\end{tikzpicture}
\end{minipage}
\begin{minipage}{0.32\linewidth}	
	\centering
	\begin{tikzpicture}
	\begin{feynman}
	\vertex (i1) {\(g\)};
	\vertex[right=1 cm of i1] (a);
	\vertex[below right=1.5 cm of a] (b);
	\vertex[right=0.8 cm of b, dot] (d) {};
	\vertex[below right=0.8 cm of d, dot] (e) {};
	\vertex[below right=0.7 cm of e] (f2) {\(H\)};
	\vertex[above right=1.4 cm of d] (f1) {\(H\)};
	
	\vertex[below=2 cm of i1] (i2) {\(g\)};
	\vertex[right=1 cm of i2] (c);
	
	\diagram* {
		(i1) -- [gluon] (a) -- [anti fermion] (b) -- [scalar] (d)[dot] -- [scalar, half left] (e)[dot] -- [scalar, half left] (d),
		(i2) -- [gluon] (c) -- [fermion] (b),
		(c) -- [anti fermion] (a),
		(d) -- [scalar] (f1),
		(e) -- [scalar] (f2)};
	\node at (2.2,-2.6) {};
	\end{feynman}
	\end{tikzpicture}
\end{minipage}
\begin{minipage}{0.32\linewidth}
	\centering
	\begin{tikzpicture}
	\begin{feynman}
	\vertex (i1) {\(g\)};
	\vertex[right=1 cm of i1] (a);
	\vertex[below right=1.5 cm of a] (b);
	\vertex[right=0.3 cm of b, dot] (d) {};
	\vertex[right=1 cm of d, dot] (e) {};
	\vertex[below right=1.4 cm of e] (f2) {\(H\)};
	\vertex[above right=1.4 cm of e] (f1) {\(H\)};
	
	\vertex[below=2 cm of i1] (i2) {\(g\)};
	\vertex[right=1 cm of i2] (c);
	
	\diagram* {
		(i1) -- [gluon] (a) -- [anti fermion] (b) -- [scalar] (d)[dot] -- [scalar, half left] (e)[dot] -- [scalar, half left] (d),
		(i2) -- [gluon] (c) -- [fermion] (b),
		(c) -- [anti fermion] (a),
		(e) -- [scalar] (f1),
		(e) -- [scalar] (f2)};
	\node at (2.5,-2.6) {};
	\end{feynman}
	\end{tikzpicture}
\end{minipage}
\begin{minipage}{0.32\linewidth}
	\centering
	\begin{tikzpicture}
	\begin{feynman}
	\vertex (i1) {\(g\)};
	\vertex[right=1 cm of i1] (a);
	\vertex[below right=1.5 cm of a] (b);
	\vertex[right=0.3 cm of b, dot] (d) {};
	\vertex[right=1 cm of d, dot] (e) {};
	\vertex[right=0.3 cm of e, dot] (f) {};
	\vertex[below right=1.4 cm of f] (f2) {\(H\)};
	\vertex[above right=1.4 cm of f] (f1) {\(H\)};
	
	\vertex[below=2 cm of i1] (i2) {\(g\)};
	\vertex[right=1 cm of i2] (c);
	
	\diagram* {
		(i1) -- [gluon] (a) -- [anti fermion] (b) -- [scalar] (d)[dot] -- [scalar, half left] (e)[dot] -- [scalar, half left] (d),
		(i2) -- [gluon] (c) -- [fermion] (b),
		(c) -- [anti fermion] (a),
		(e) -- [scalar] (f)[dot] --[scalar] (f1),
		(f) -- [scalar] (f2)};
	\node at (2.7,-2.6) {};
	\end{feynman}
	\end{tikzpicture}
\end{minipage}
\begin{minipage}{0.32\linewidth}
		\centering
		\begin{tikzpicture}
		\begin{feynman}
		\vertex (i1) {\(g\)};
		\vertex[right=1 cm of i1] (a);
		\vertex[right=1 cm of a] (b);
		\vertex[below right=1.4 cm of b, dot] (e) {};
		\vertex[above right=1.4 cm of e] (f1) {\(H\)};
		\vertex[below right=1.4 cm of e] (f2) {\(H\)};
		
		\vertex[below=2 cm of i1] (i2) {\(g\)};
		\vertex[right=1 cm of i2] (c);
		\vertex[right=1 cm of c] (d);
		
		\diagram* {
			(i1) -- [gluon] (a) -- [anti fermion] (b) -- [scalar] (e)[dot] -- [scalar] (f1),
			(i2) -- [gluon] (c) -- [fermion] (d) -- [scalar] (e) -- [scalar] (f2),
			(a) -- [fermion] (c),
			(b) -- [anti fermion] (d)};
		\node at (2.2,-2.6) {};
		\end{feynman}
		\end{tikzpicture}
	\end{minipage}
 \begin{minipage}{0.32\linewidth}	
 	\centering
 	\begin{tikzpicture}
 	\begin{feynman}
 	\vertex (i1) {\(g\)};
 	\vertex[right=1 cm of i1] (a);
 	\vertex[right=1 cm of a] (b);
 	\vertex[below right=1.4 cm of b, dot] (e) {};
 	\vertex[right=0.5 cm of e, dot] (f) {};
 	\vertex[above right=1.4 cm of f] (f1) {\(H\)};
 	\vertex[below right=1.4 cm of f] (f2) {\(H\)};
 	
 	\vertex[below=2 cm of i1] (i2) {\(g\)};
 	\vertex[right=1 cm of i2] (c);
 	\vertex[right=1 cm of c] (d);
 	
 	\diagram* {
 		(i1) -- [gluon] (a) -- [anti fermion] (b) -- [scalar] (e)[dot] -- [scalar] (f)[dot] -- [scalar] (f1),
 		(i2) -- [gluon] (c) -- [fermion] (d) -- [scalar] (e),
 		(a) -- [fermion] (c),
 		(b) -- [anti fermion] (d),
 		(f) -- [scalar] (f2)};
 	\node at (2.4,-2.6) {};
 	\end{feynman}
 	\end{tikzpicture}
 \end{minipage}
		\begin{minipage}{0.32\linewidth}
		\centering
		\begin{tikzpicture}
		\begin{feynman}
		\vertex (i1) {\(g\)};
		\vertex[right=1 cm of i1] (a);
		\vertex[right=1 cm of a] (b);
		\vertex[right=0.8 cm of b, dot] (e) {};
		\vertex[right=1 cm of e] (f1) {\(H\)};
		
		\vertex[below=2 cm of i1] (i2) {\(g\)};
		\vertex[right=1 cm of i2] (c);
		\vertex[right=1 cm of c] (d);
		\vertex[right=0.8 cm of d, dot] (f) {};
		\vertex[right=1 cm of f] (f2) {\(H\)};
		
		\diagram* {
			(i1) -- [gluon] (a) -- [anti fermion] (b) -- [scalar] (e)[dot] -- [scalar] (f1),
			(i2) -- [gluon] (c) -- [anti fermion] (d) -- [scalar] (f)[dot] -- [scalar] (f2),
			(a) -- [fermion] (d),
			(b) -- [anti fermion] (c),
			(e) -- [scalar] (f)};
		\node at (2.2,-2.6) {};
		\end{feynman}
		\end{tikzpicture}
	\end{minipage}
	\begin{minipage}{0.32\linewidth}
		\centering
		\begin{tikzpicture}
		\begin{feynman}
		\vertex (i1) {\(g\)};
		\vertex[right=1 cm of i1] (a);
		\vertex[right=1 cm of a] (b);
		\vertex[right=0.8 cm of b, dot] (e) {};
		\vertex[right=1 cm of e] (f1) {\(H\)};
		
		\vertex[below=2 cm of i1] (i2) {\(g\)};
		\vertex[right=1 cm of i2] (c);
		\vertex[right=1 cm of c] (d);
		\vertex[right=0.8 cm of d, dot] (f) {};
		\vertex[right=1 cm of f] (f2) {\(H\)};
		
		\diagram* {
			(i1) -- [gluon] (a) -- [anti fermion] (b) -- [scalar] (e)[dot] -- [scalar] (f1),
			(i2) -- [gluon] (c) -- [fermion] (d) -- [scalar] (f)[dot] -- [scalar] (f2),
			(a) -- [fermion] (c),
			(b) -- [anti fermion] (d),
			(e) -- [scalar] (f)};
		\node at (2.2,-2.6) {};
		\end{feynman}
		\end{tikzpicture}
	\end{minipage}
	\begin{minipage}{0.32\linewidth}
		\centering
		\begin{tikzpicture}
		\begin{feynman}
		\vertex (i1) {\(g\)};
		\vertex[right=1 cm of i1] (a);
		\vertex[right=1 cm of a] (b);
		\vertex[right=0.8 cm of b, dot] (e) {};
		\vertex[right=1 cm of e] (f1) {\(H\)};
		
		\vertex[below=2 cm of i1] (i2) {\(g\)};
		\vertex[right=1 cm of i2] (c);
		\vertex[right=1 cm of c] (d);
		\vertex[right=0.8 cm of d, dot] (f) {};
		\vertex[right=1 cm of f] (f2) {\(H\)};
		
		\diagram* {
			(i1) -- [gluon] (a) -- [anti fermion] (b) -- [scalar] (e)[dot] -- [scalar] (f2),
			(i2) -- [gluon] (c) -- [fermion] (d) -- [scalar] (f)[dot] -- [scalar] (f1),
			(a) -- [fermion] (c),
			(b) -- [anti fermion] (d),
			(e) -- [scalar] (f)};
		\node at (2.2,-2.6) {};
		\end{feynman}
		\end{tikzpicture}
	\end{minipage}
	\begin{minipage}{0.32\linewidth}	
		\centering
		\begin{tikzpicture}
		\begin{feynman}
		\vertex (i1) {\(g\)};
		\vertex[right=1 cm of i1] (a);
		\vertex[right=1 cm of a] (b);
		\vertex[below right=1.4 cm of b, dot] (e) {};
		\vertex[above right=1.4 cm of e] (f1) {\(H\)};
		\vertex[below right=1.4 cm of e] (f2) {\(H\)};
		
		\vertex[below=2 cm of i1] (i2) {\(g\)};
		\vertex[right=1 cm of i2] (c);
		\vertex[right=1 cm of c] (d);
		
		\diagram* {
			(i1) -- [gluon] (a) -- [anti fermion] (b) -- [scalar] (e)[dot] -- [scalar] (f1),
			(i2) -- [gluon] (c) -- [anti fermion] (d) -- [scalar] (e) -- [scalar] (f2),
			(a) -- [fermion] (d),
			(b) -- [anti fermion] (c)};
		\node at (2.2,-2.6) {};
		\end{feynman}
		\end{tikzpicture}
	\end{minipage}
	\begin{minipage}{0.32\linewidth}	
		\centering
		\begin{tikzpicture}
		\begin{feynman}
		\vertex (i1) {\(g\)};
		\vertex[right=1 cm of i1] (a);
		\vertex[right=1 cm of a] (b);
		\vertex[below right=1.4 cm of b, dot] (e) {};
		\vertex[right=0.5 cm of e, dot] (f) {};
		\vertex[above right=1.4 cm of f] (f1) {\(H\)};
		\vertex[below right=1.4 cm of f] (f2) {\(H\)};
		
		\vertex[below=2 cm of i1] (i2) {\(g\)};
		\vertex[right=1 cm of i2] (c);
		\vertex[right=1 cm of c] (d);
		
		\diagram* {
			(i1) -- [gluon] (a) -- [anti fermion] (b) -- [scalar] (e)[dot] -- [scalar] (f)[dot] -- [scalar] (f1),
			(i2) -- [gluon] (c) -- [anti fermion] (d) -- [scalar] (e),
			(f) -- [scalar] (f2),
			(a) -- [fermion] (d),
			(b) -- [anti fermion] (c)};
		\node at (2.5,-2.6) {};
		\end{feynman}
		\end{tikzpicture}
	\end{minipage}
	\caption{Feynman diagrams contributing to $gg \to HH$ that contain at least two Higgs trilinear self-couplings or a single Higgs quartic self-coupling. The modified Higgs self-interactions are indicated by black dots. The diagrams with reversed fermion flow are omitted.}
	\label{figNLO_HH}
\end{figure}

The NLO~EW~corrections to double-Higgs production involving $\kappa_3$ and $\kappa_4$ can be computed within the~HEFT framework, using the EW~chiral Lagrangian, which can be regarded as an extension of the~$\kappa$-framework to a fully consistent quantum field theory. At the LO the EW~chiral Lagrangian is given by~\cite{Buchalla:2013rka}
\beq \label{eq:L2EchL}
 {\cal L}_2 = \frac{1}{2} \hspace{0.25mm} (\partial_\mu H) (\partial^\mu H) - T - V +\frac{v^2}{4} \left(1 + 2 c_1 \hspace{0.25mm} \frac{H}{v} + c_2 \left(\frac{H}{v}\right)^2 + \cdots \right) {\rm Tr} \left [D_{\mu}U^{\dagger} D^{\mu} U \right ] + \cdots\,, 
\eeq
where we have written the Higgs potential $V$ and the interaction terms between the Higgs boson and the EW~gauge bosons explicitly, assuming that the fermionic sector remains the same as in the SM. The~matrix $U$ parameterizes the Goldstone bosons $\pi^k$ nonlinearly,
\beq \label{eq:U}
U = e^{\frac{i \pi^k \sigma^k}{v}} \,,
\eeq
where $\sigma^k$ are the Pauli matrices and $k = 1,2,3$. The covariant derivative acting on $U$ is defined~as
\beq \label{eq:covariant}
D_{\mu} U = \partial_{\mu} U + i g_2 W^k_{\mu} \hspace{0.25mm} \frac{\sigma^k}{2} U - i g_1 B_{\mu} U \hspace{0.25mm} \frac{\sigma^3}{2} \,.
\eeq
Here, $W^k_{\mu}$ and $B_\mu$ denote the $SU(2)_L$ and $U(1)_Y$ gauge fields, respectively. The factor $v^2/4$ in~Eq.~(\ref{eq:L2EchL}) is chosen to reproduce the correct masses of the EW~gauge bosons, while the coefficients $c_1$ and $c_2$ parameterize the couplings between the EW~gauge bosons and one or two Higgs bosons, respectively. Besides the Higgs potential $V$ given in~Eq.~(\ref{eq:V}), the Lagrangian~(\ref{eq:L2EchL}) also contains the Higgs tadpole~term
\beq \label{eq:T}
T = \left( m_H^2 - 2 \lambda v^2 \right) v H \,. 
\eeq
This term vanishes at tree level due to $m_H^2 = 2 \lambda v^2$ and has therefore been omitted in many studies based on the EW~chiral Lagrangian. However, it is required once the renormalization of the Higgs field is considered. Note also that the SM Lagrangian~(\ref{eq:lag2}) is recovered from Eqs.~(\ref{eq:V}), (\ref{eq:L2EchL}), and~(\ref{eq:T}) when $c_1 = c_2 = \kappa_3 = \kappa_4 = 1$ and all $\kappa_n = 0$ for $n \ge 5$.

In the case of double-Higgs production in ggF, there are two types of NLO~EW~corrections that modify the dependence of the differential cross section on the Higgs self-couplings. The~first type arises from diagrams with at most one triple-Higgs vertex. The diagrams which contain a single Higgs trilinear self-coupling and three top-quark Yukawa couplings are shown in~Figure~\ref{figNLO_HY}. These contributions are ultraviolet~(UV) finite for the $gg \to HH$ amplitude, as the Higgs trilinear self-coupling does not mix with the top-quark Yukawa coupling at the one-loop order. This mixing first appears at the two-loop level, giving rise to a non-zero anomalous dimension~\cite{Gorbahn:2016uoy}. The interference of these two-loop diagrams with the LO amplitudes yields the following additive~corrections
\beq \label{eq:typeI}
\begin{split}
\sigma^{13\,{\rm TeV}}_{\rm ggF,EW_I} & = \left ( -0.176 \hspace{0.25mm} \kappa_{3}^2+0.572 \hspace{0.25mm} \kappa_{3} \right ) {\rm fb} \,, \\[2mm] 
\sigma^{13.6\,{\rm TeV}}_{\rm ggF,EW_I} & = \left ( -0.194 \hspace{0.25mm} \kappa_{3}^2+0.632 \hspace{0.25mm} \kappa_{3} \right ) {\rm fb} \,, \\[2mm] 
\sigma^{14\,{\rm TeV}}_{\rm ggF,EW_I} & = \left ( -0.207 \hspace{0.25mm} \kappa_{3}^2+0.674 \hspace{0.25mm} \kappa_{3} \right ) {\rm fb} \,, 
\end{split}
\eeq
which exhibit a mild dependence on the collider energy~$\sqrt{s}$. Compared to the NLO~QCD corrections~(\ref{eq:ggFNLO}), these NLO~EW~effects change the linear (quadratic) $\kappa_3$ dependence of the ggF double-Higgs production cross section by about $-1\%$ ($-2\%$). Comparable effects are expected from the NLO~EW~corrections involving the EW~gauge bosons, as demonstrated in the case of the SM in~Ref.~\cite{Bi:2023bnq}.

A second type of NLO~EW~corrections to double-Higgs production arises from diagrams containing at least two triple-Higgs vertices or a Higgs quartic self-coupling. For the ggF channel, the relevant two-loop diagrams are shown in~Figure~\ref{figNLO_HH}. Given the currently weak bounds on $\kappa_3$ and $\kappa_4$ in~Eqs.~(\ref{eq:kappa3bounds}) and~(\ref{eq:kappa4bound}), this type of NLO~EW~contribution can be phenomenologically much larger than the corrections in~Eqs.~(\ref{eq:typeI}). The~calculation of this second type of NLO~EW~corrections requires renormalization of the Higgs sector within the HEFT, as the diagrams in~Figure~\ref{figNLO_HH} contain UV divergences. To achieve this, we introduce renormalization constants via redefinitions of the fields and parameters:
\beq \label{eq:renormalization}
H=Z_H^{1/2} H_R \, \quad 
m_{H}^2 = Z_{m^2} \hspace{0.25mm} m_{H, R}^{2} \, \quad 
v=Z_v v_R, \quad \lambda = Z_{\lambda} \hspace{0.25mm} \lambda_R\, \quad 
\kappa_3 = Z_{\kappa_3}\kappa_{3,R} \,, \quad 
\kappa_4 = Z_{\kappa_4}\kappa_{4,R} \,,
\eeq
where the subscript $R$ denotes renormalized quantities. Since the Higgs quartic self-coupling enters only at the two-loop level, the coupling modifier $\kappa_4$ does not require renormalization, implying that the corresponding $Z$ factor can be set to one,~i.e., $Z_{\kappa_4} = 1$. 

Inserting the renormalized quantities~(\ref{eq:renormalization}) back into the Lagrangian~(\ref{eq:L2EchL}) yields the following counterterms for the one-, two-, and three-point Higgs vertices:
\begin{align}
 \feynmandiagram[horizontal=a to b]{
	 a --[scalar] b[crossed dot]};
 \qquad &= -2i\lambda v^3 \left (\delta Z_{m^2}-2\delta Z_v -\delta Z_{\lambda} \right ) \,, \nonumber \\[4mm]
 \feynmandiagram[horizontal=a to c]{
	a --[scalar]b[crossed dot]--[scalar]c};
 \qquad &= i \left [p^2 \delta Z_H - \left (\delta Z_{m^2}+\delta Z_H \right )m_H^2 \right ]\,, \label{eq:counterterms} \\[2mm]
 &\hspace{-2.5cm}	
 \begin{tikzpicture}
	\begin{feynman}
	\vertex (a);
	\vertex [right= 1 cm of a, crossed dot] (b){};
	\vertex [above right= 1.2 cm of b](c);
	\vertex [below right= 1.2 cm of b](d);
	\diagram*{
		(a) --[scalar](b)[cross dot] --[scalar](c),
		(b) --[scalar](d)};
 \node at (5.2,0) {$~~~~~~~= -6i\kappa_3 \lambda v \left( \displaystyle\frac{3}{2} \hspace{0.5mm} \delta Z_H + \delta Z_{\lambda} +\delta Z_v +\delta Z_{\kappa_3} \right)$\,.};
	\end{feynman} \nonumber 
	\end{tikzpicture}
\end{align}
Here, we have used the notation $\delta Z = Z - 1$. The renormalization conditions are chosen as follows:
\begin{itemize}
\item The Higgs tadpole contribution is renormalized to zero.
\item The Higgs mass and wave function are renormalized in the on-shell scheme.
\item The coupling modifier $\kappa_3$ is renormalized in the $\overline{\rm MS}$ scheme.
\end{itemize}
Since the renormalization of the $W$- and $Z$-boson masses, as well as the electric charge, does not involve the Higgs trilinear self-coupling at the one-loop level, we can leave the VEV~$v$ unrenormalized by setting $\delta Z_v = 0$ without loss of generality. Making use of the above renormalization conditions, the $Z$ factors are uniquely determined. For instance, for the $\overline{\rm MS}$ counterterm of $\kappa_3$, we obtain
\beq \label{eq:kappa3ct}
\delta Z_{\kappa_3}=\frac{3\lambda}{16\pi^2}\frac{1}{\epsilon}\left(\kappa_3 +2\kappa_4 -3\kappa_3^2 \right) \,,
\eeq
 where the UV pole $1/\epsilon$ with $\epsilon = (4-d)/2$ arises from dimensional regularization in $d$ dimensions. The~inclusion of the counterterm contributions~(\ref{eq:counterterms}) removes all UV divergences present in the bare two-loop amplitudes.

In the following, we present results for the second type of NLO~EW~contributions for both ggF and VBF double-Higgs production, which constitute the two dominant double-Higgs production mechanisms at the LHC. Since the two-loop $gg \to HH$ diagrams involve multiple scales, obtaining analytical results is currently beyond the state-of-the-art. We therefore adopt a numerical approach using the~\texttt{AMFlow}~package~\cite{Liu:2017jxz,Liu:2022chg}. In contrast, the loop diagrams in the VBF channel are computed analytically and implemented into the \texttt{proVBFHH} program~\cite{Cacciari:2015jma,Dreyer:2016oyx,Dreyer:2018rfu,Dreyer:2020urf} to perform MC integrations. To obtain numerical predictions, we use the input parameters
\begin{gather} 
v = 246.2 \, {\rm GeV} \,, \quad 
m_H = 125 \, {\rm GeV} \,, \quad 
m_t = 173 \, {\rm GeV} \,, \nonumber \\[-2mm] \label{eq:input} \\[-2mm]
m_W = 80.379 \, {\rm GeV} \,, \quad 
m_Z = 91.1876 \, {\rm GeV} \,, \nonumber
\end{gather}
and employ the \texttt{PDF4LHC15\_NLO} PDF set together with the associated strong coupling~$\alpha_s$. In the ggF double-Higgs production cross-section calculation, the renormalization and factorization scales are chosen as $\mu_R = \mu_F = m_{HH}/2$, while in the VBF channel we use $\mu_R = \mu_F = \sqrt{-q^2}$, where~$q^2$ denotes the momentum squared transferred in the $t$-channel exchange between the two quark lines.

\begin{figure}[t!]
\centering
\includegraphics[width=7cm]{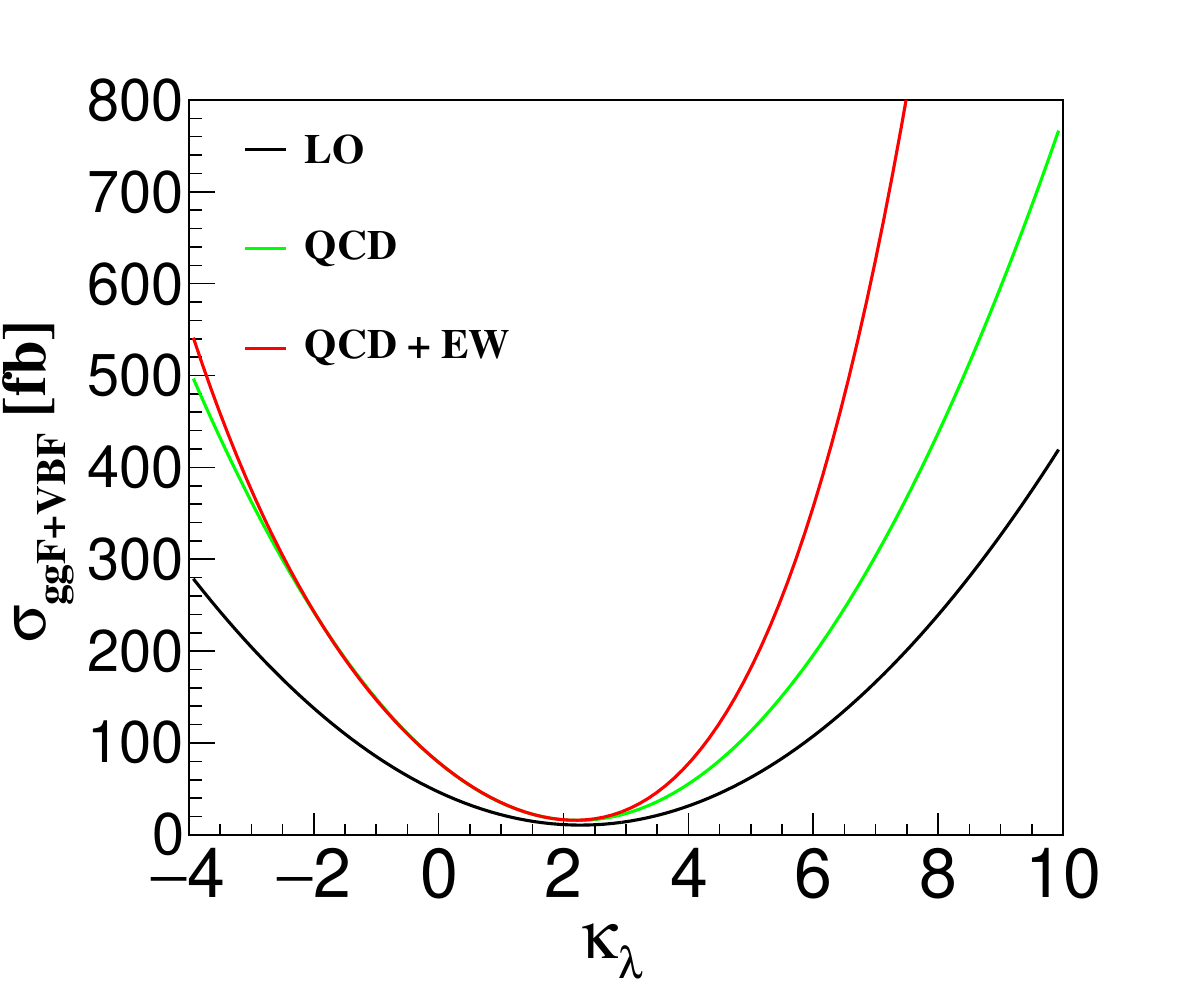} \qquad 
\includegraphics[width=7cm]{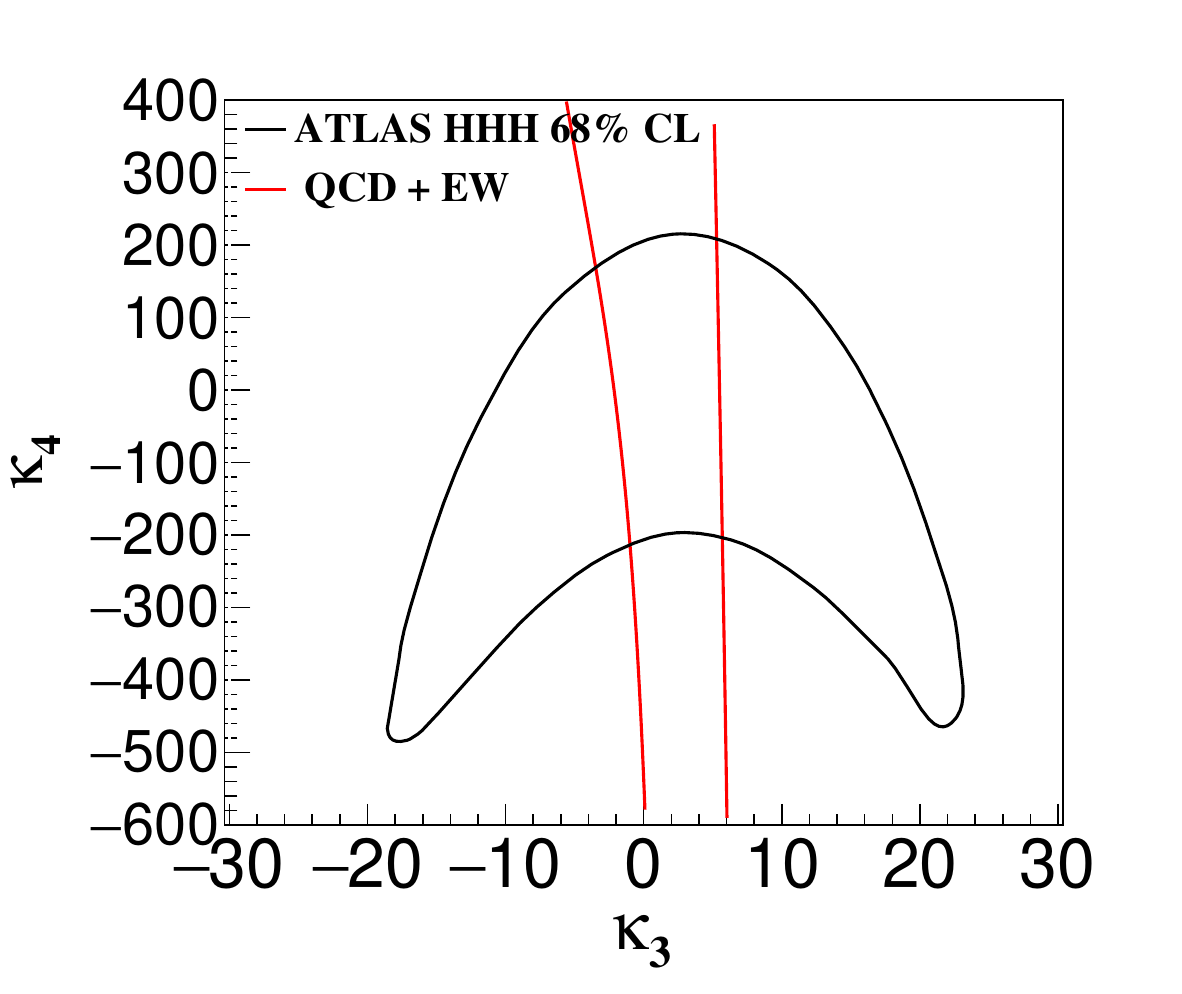}
\vspace{2mm} 
\caption{Left: Cross sections for double-Higgs production in the combined ggF and VBF channels at the LHC with $\sqrt{s} = 14 \, {\rm TeV}$ as a function of $\kappa_\lambda = \kappa_3 = \kappa_4$. The black line shows the LO prediction, the green line includes QCD corrections, and the red line incorporates NLO QCD and NLO EW in ggF and N$^3$LO QCD and NLO EW in VBF, collectively referred to as QCD+EW. Right: Constraints in the $\kappa_3\hspace{0.25mm}-\hspace{0.25mm}\kappa_4$ plane from LHC~Run~2. The red contour corresponds to the preferred $68\%$~CL region derived from the HEFT calculation of the QCD+EW corrections to double-Higgs production presented here. For comparison, the $68\%$~CL region from triple-Higgs production, as obtained by the ATLAS collaboration~\cite{ATLAS:2024xcs}, is shown as a black~contour.}
\label{fig:kappa}
\end{figure}

The interference between the two-loop diagrams shown in~Figure~\ref{figNLO_HH}, including the corresponding one-loop counterterms, and the LO $gg \to HH$ amplitudes, yields the following additive contributions to the ggF double-Higgs production cross section:
\beq \label{eq:EWIIggF}
\begin{split}
\sigma^{13\,{\rm TeV}}_{\rm ggF,EW_{II}} & = \left ( 0.067\hspace{0.25mm}\kappa_{3}^4-0.105\hspace{0.25mm}\kappa_{3}^3-0.006\hspace{0.25mm}\kappa_{3}^2\kappa_{4}-0.166\hspace{0.25mm}\kappa_{3}^2+0.070\hspace{0.25mm}\kappa_{3}\kappa_{4}-0.149\kappa_{4} \right ) {\rm fb} \, ,\\[2mm]
\sigma^{13.6\,{\rm TeV}}_{\rm ggF,EW_{II}} & = \left ( 0.074\hspace{0.25mm}\kappa_{3}^4-0.114\hspace{0.25mm}\kappa_{3}^3-0.006\hspace{0.25mm}\kappa_{3}^2\kappa_{4}-0.185\hspace{0.25mm}\kappa_{3}^2+0.077\hspace{0.25mm}\kappa_{3}\kappa_{4} -0.163\hspace{0.25mm}\kappa_{4} \right ) {\rm fb} \, ,\\[2mm]
\sigma^{14\,{\rm TeV}}_{\rm ggF,EW_{II}} & = \left ( 0.078\hspace{0.25mm}\kappa_{3}^4-0.120\hspace{0.25mm}\kappa_{3}^3-0.006\hspace{0.25mm}\kappa_{3}^2\hspace{0.25mm}\kappa_{4}-0.199\hspace{0.25mm}\kappa_{3}^2+0.081\hspace{0.25mm}\kappa_{3}\kappa_{4} -0.173\hspace{0.25mm}\kappa_{4}\right ) {\rm fb} \, .
\end{split}
\eeq
The corresponding NLO~EW~corrections of the second type for VBF double-Higgs production are given~by
\beq \label{eq:EWIIVBF}
\begin{split}
\sigma_{\rm VBF,EW_{II}}^{13\,{\rm TeV}}&=\left ( 1.96\hspace{0.25mm}\kappa_{3}^4-2.35\hspace{0.25mm}\kappa_{3}^3-0.14\hspace{0.25mm}\kappa_{3}^2\kappa_{4}-1.65\hspace{0.25mm}\kappa_{3}^2 +1.26\hspace{0.25mm}\kappa_{3}\kappa_{4}-1.93\hspace{0.25mm}\kappa_{4} \right )\cdot 10^{-2} \, {\rm fb} \, , \\[2mm]
\sigma_{\rm VBF,EW_{II}}^{13.6\,{\rm TeV}}&=\left ( 2.11\hspace{0.25mm}\kappa_{3}^4-2.00\hspace{0.25mm}\kappa_{3}^3-0.44\hspace{0.25mm}\kappa_{3}^2\kappa_{4}-3.76\hspace{0.25mm}\kappa_{3}^2 +3.35\hspace{0.25mm}\kappa_{3}\kappa_{4}-3.56\hspace{0.25mm}\kappa_{4} \right )\cdot 10^{-2} \, {\rm fb} \, , \\[2mm]
\sigma_{\rm VBF,EW_{II}}^{14\,{\rm TeV}}&=\left ( 2.27\hspace{0.25mm}\kappa_{3}^4-2.65\hspace{0.25mm}\kappa_{3}^3-0.21\hspace{0.25mm}\kappa_{3}^2\kappa_{4}-2.16\hspace{0.25mm}\kappa_{3}^2+1.74\hspace{0.25mm}\kappa_{3}\kappa_{4}-2.46\hspace{0.25mm}\kappa_{4} \right )\cdot 10^{-2} \, {\rm fb} \, . 
\end{split}
\eeq
The results in~Eqs.~(\ref{eq:EWIIVBF}) are obtained under the assumption $c_1 = c_2 = 1$, see~Eq.~(\ref{eq:L2EchL}). Comparing the~ggF and VBF results, one observes that although the cross-section modifications in the VBF channel are generally smaller than in the ggF process, the dependence on $\kappa_3$ and~$\kappa_4$ differs markedly between the two channels. In addition, the $\sqrt{s}$ dependence of the cross sections is more pronounced for VBF than for ggF double-Higgs production. A combination of both channels therefore leads to an improved sensitivity to $\kappa_3$ and $\kappa_4$. 

Combining the results presented in Eqs.~(\ref{eq:ggFNLO}), (\ref{eq:VBF}), (\ref{eq:typeI}), (\ref{eq:EWIIggF}), and (\ref{eq:EWIIVBF}), we obtain the following expressions for the inclusive double-Higgs production signal strengths in the HEFT:
\beq \label{eq:muhhsHEFT}
\begin{split}
\mu_{2h, \mathrm{HEFT}}^{13 \, {\rm TeV}} & = 2.22 - 1.55 \hspace{0.25mm} \kappa_3 - 5.69 \cdot 10^{-3} \hspace{0.25mm} \kappa_4 + 3.39 \cdot 10^{-1} \hspace{0.25mm} \kappa_3^2 + 2.79 \cdot 10^{-3} \hspace{0.25mm} \kappa_3 \hspace{0.25mm} \kappa_4 \\[1mm]
& \phantom{xx} - 2.50 \cdot 10^{-4} \hspace{0.25mm} \kappa_3^2 \hspace{0.25mm} \kappa_4 - 4.34 \cdot 10^{-3} \hspace{0.25mm} \kappa_3^3 + 2.93 \cdot 10^{-3} \hspace{0.25mm} \kappa_3^4 \,, \\[2mm]
\mu_{2h, \mathrm{HEFT}}^{13.6 \, {\rm TeV}} & = 2.22 - 1.55 \hspace{0.25mm} \kappa_3 - 6.06 \cdot 10^{-3} \hspace{0.25mm} \kappa_4 + 3.38 \cdot 10^{-1} \hspace{0.25mm} \kappa_3^2 + 3.37 \cdot 10^{-3} \hspace{0.25mm} \kappa_3 \hspace{0.25mm} \kappa_4 \\[1mm]
& \phantom{xx} - 3.17 \cdot 10^{-4} \hspace{0.25mm} \kappa_3^2 \hspace{0.25mm} \kappa_4 - 4.09 \cdot 10^{-3} \hspace{0.25mm} \kappa_3^3 + 2.90 \cdot 10^{-3} \hspace{0.25mm} \kappa_3^4 \,, \\[2mm]
\mu_{2h, \mathrm{HEFT}}^{14 \, {\rm TeV}} & = 2.22 - 1.55 \hspace{0.25mm} \kappa_3 - 5.63 \cdot 10^{-3} \hspace{0.25mm} \kappa_4 + 3.39 \cdot 10^{-1} \hspace{0.25mm} \kappa_3^2 + 2.80 \cdot 10^{-3} \hspace{0.25mm} \kappa_3 \hspace{0.25mm} \kappa_4 \\[1mm]
& \phantom{xx} - 2.31 \cdot 10^{-4} \hspace{0.25mm} \kappa_3^2 \hspace{0.25mm} \kappa_4 - 4.17 \cdot 10^{-3} \hspace{0.25mm} \kappa_3^3 + 2.87 \cdot 10^{-3} \hspace{0.25mm} \kappa_3^4 \,.
\end{split}
\eeq
Notice that, unlike Eqs.~(\ref{eq:muhhs}), which consider only ggF double-Higgs production corrections, these expressions include contributions from both the ggF and VBF channels.

\begin{figure}[t!]
\centering
\includegraphics[width=4.9cm]{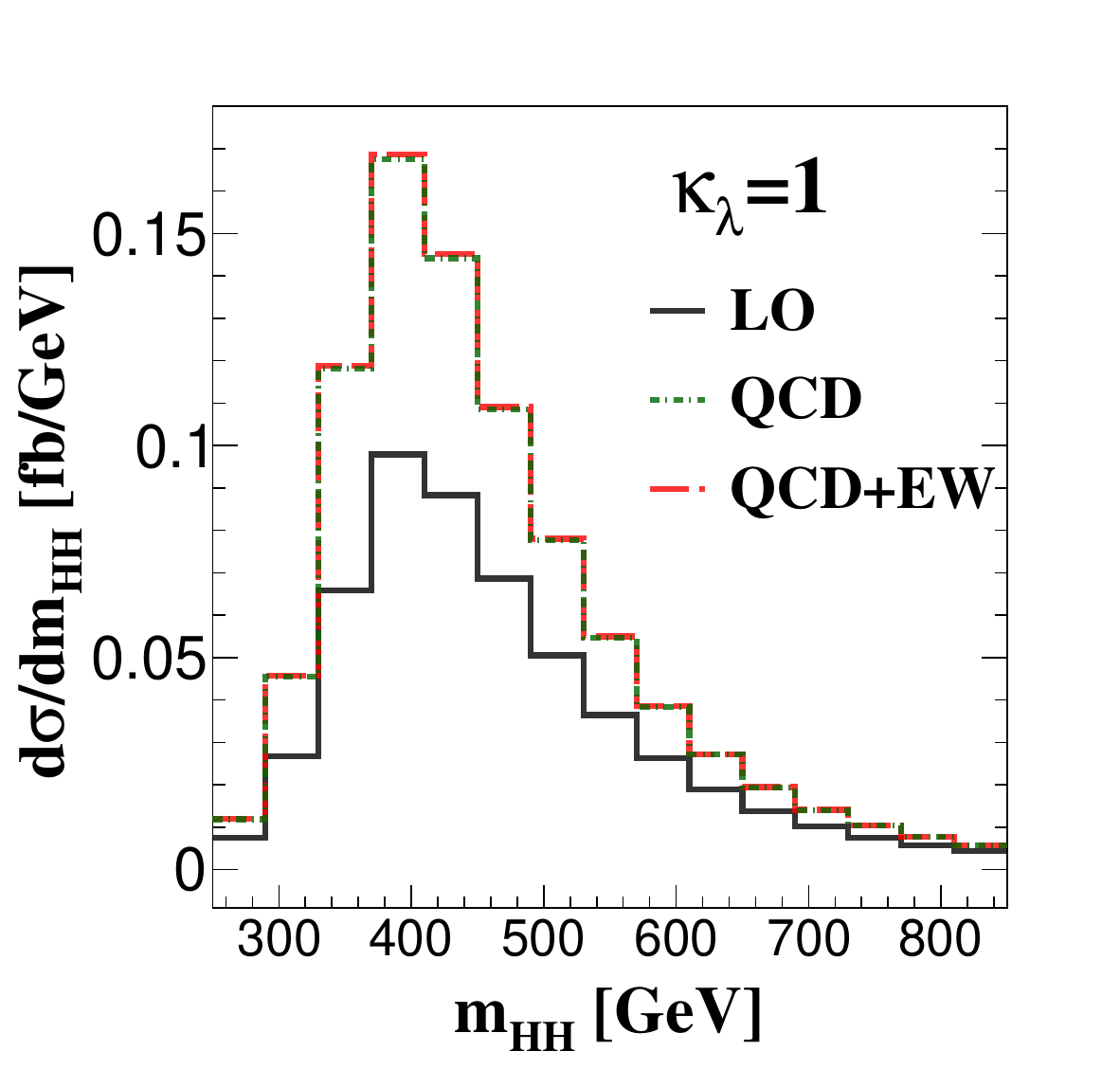}
\includegraphics[width=4.9cm]{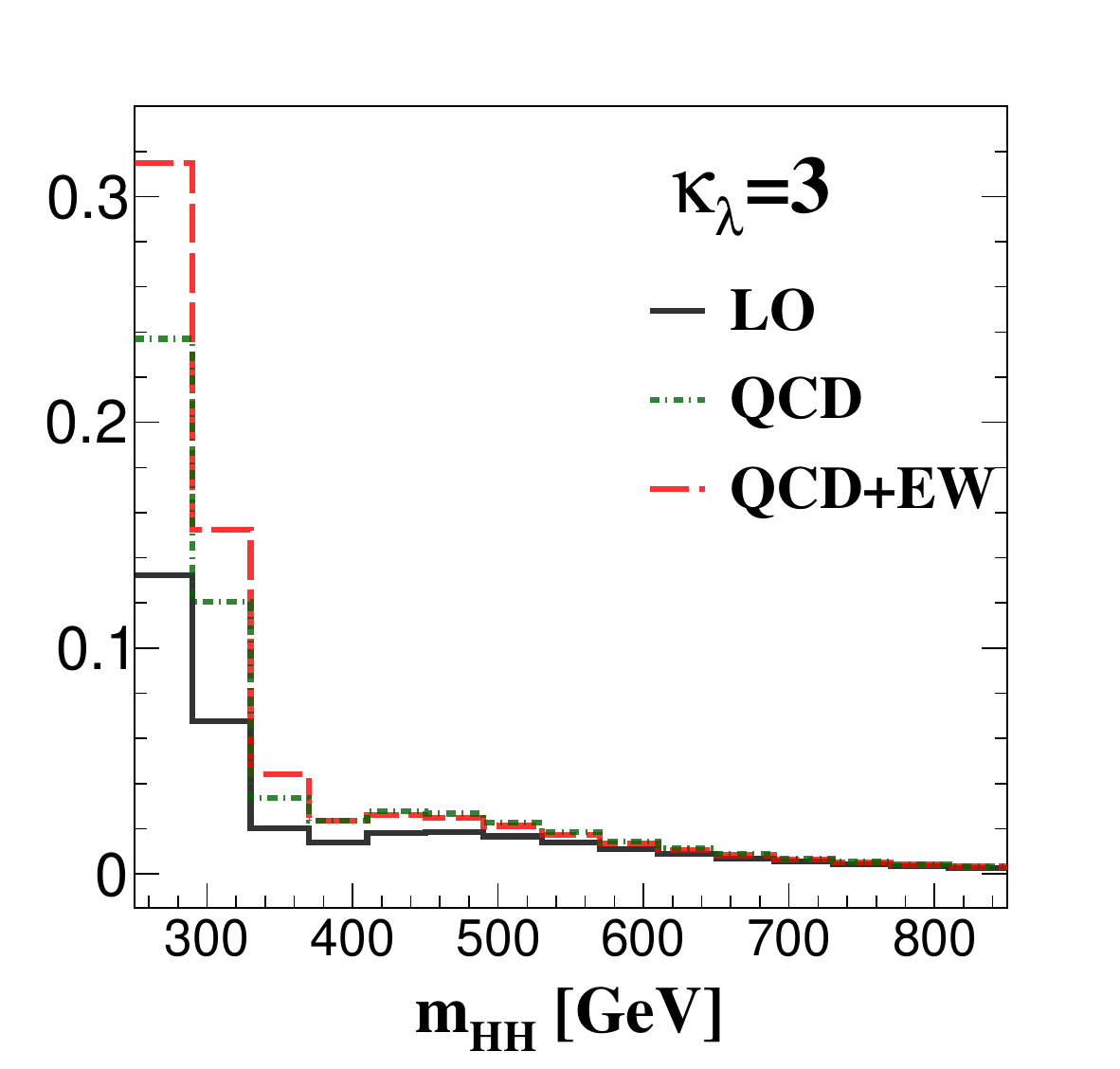}
\includegraphics[width=4.9cm]{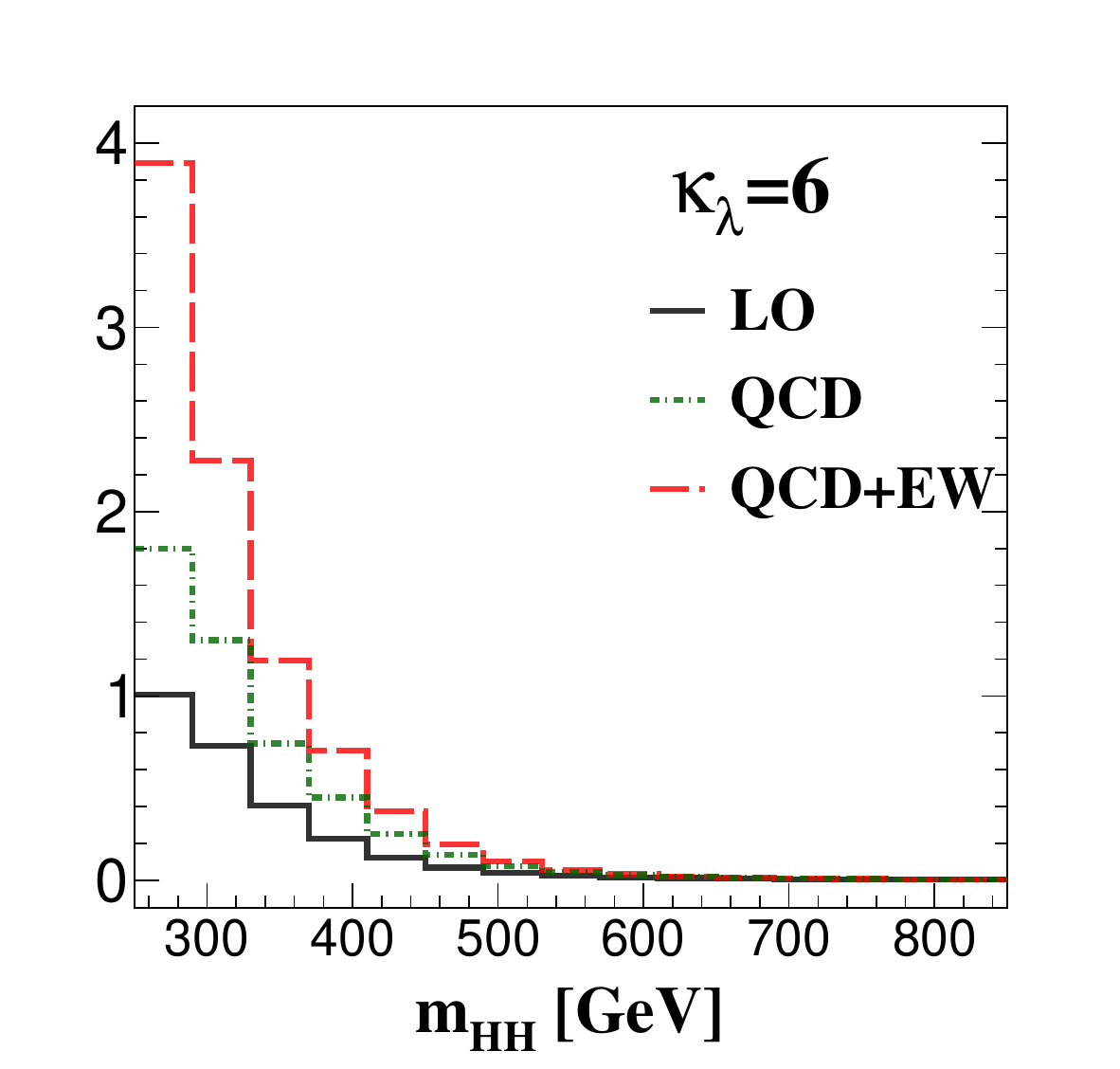}	\\
\includegraphics[width=4.9cm]{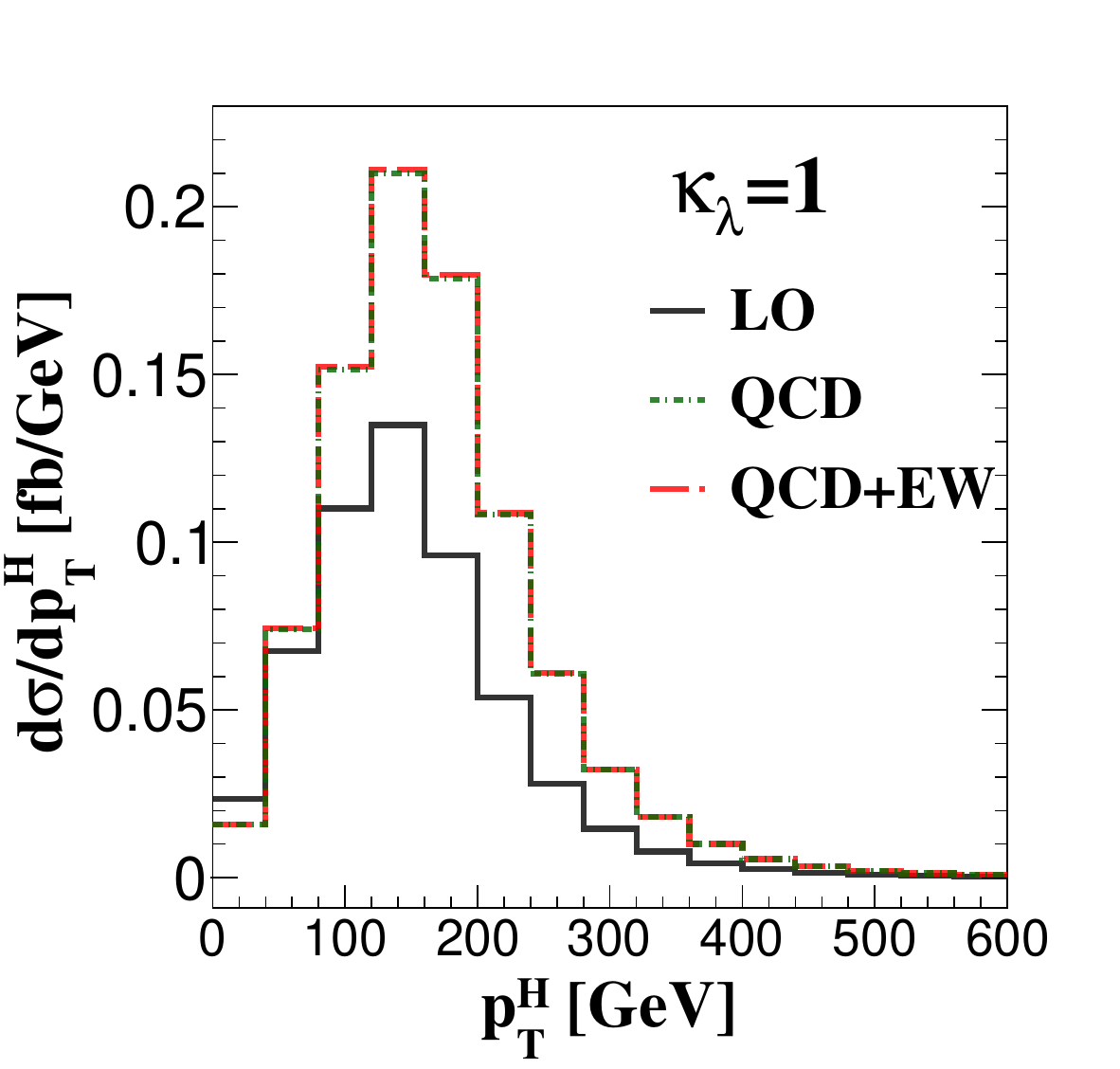}
\includegraphics[width=4.9cm]{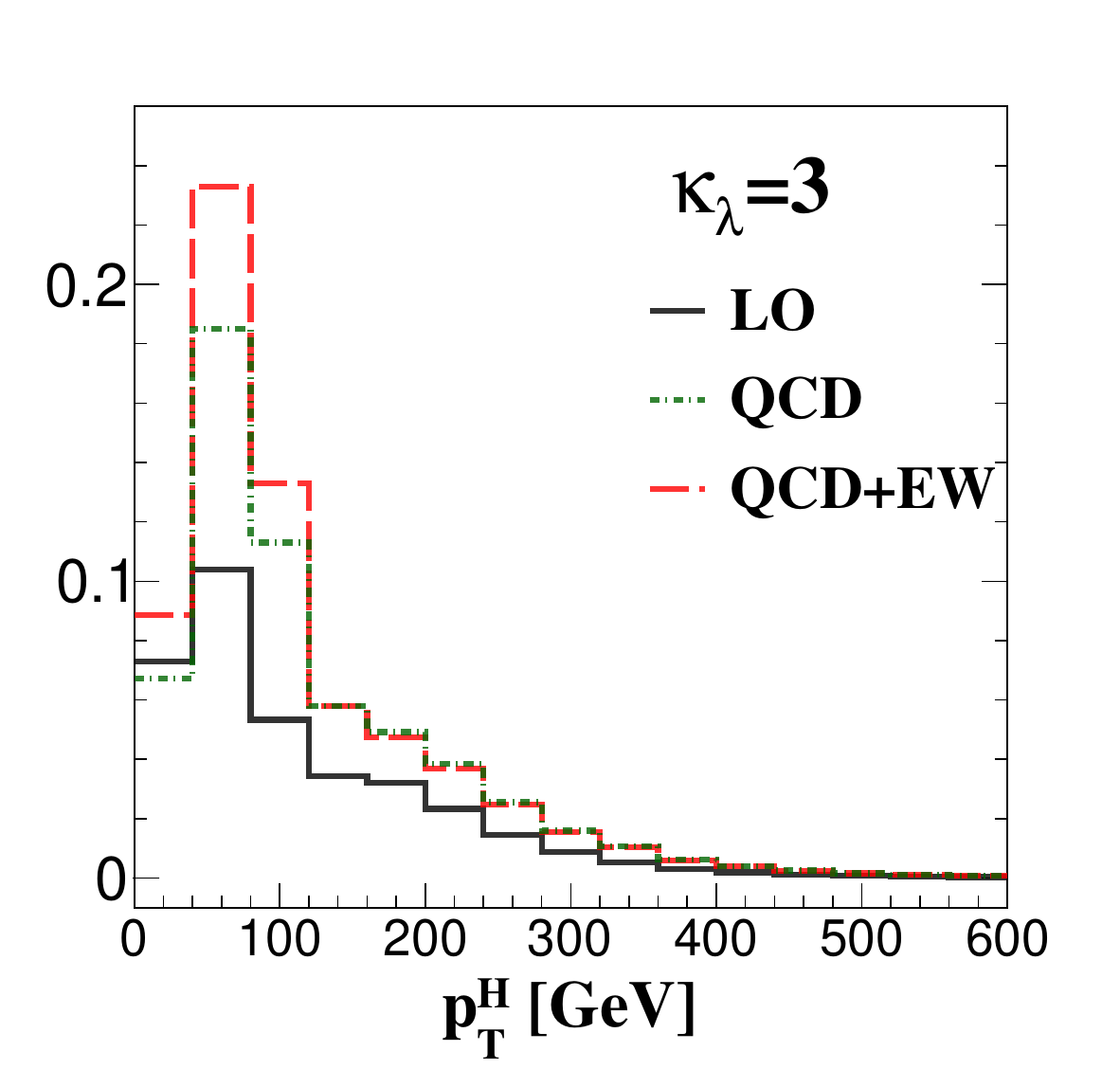}
\includegraphics[width=4.9cm]{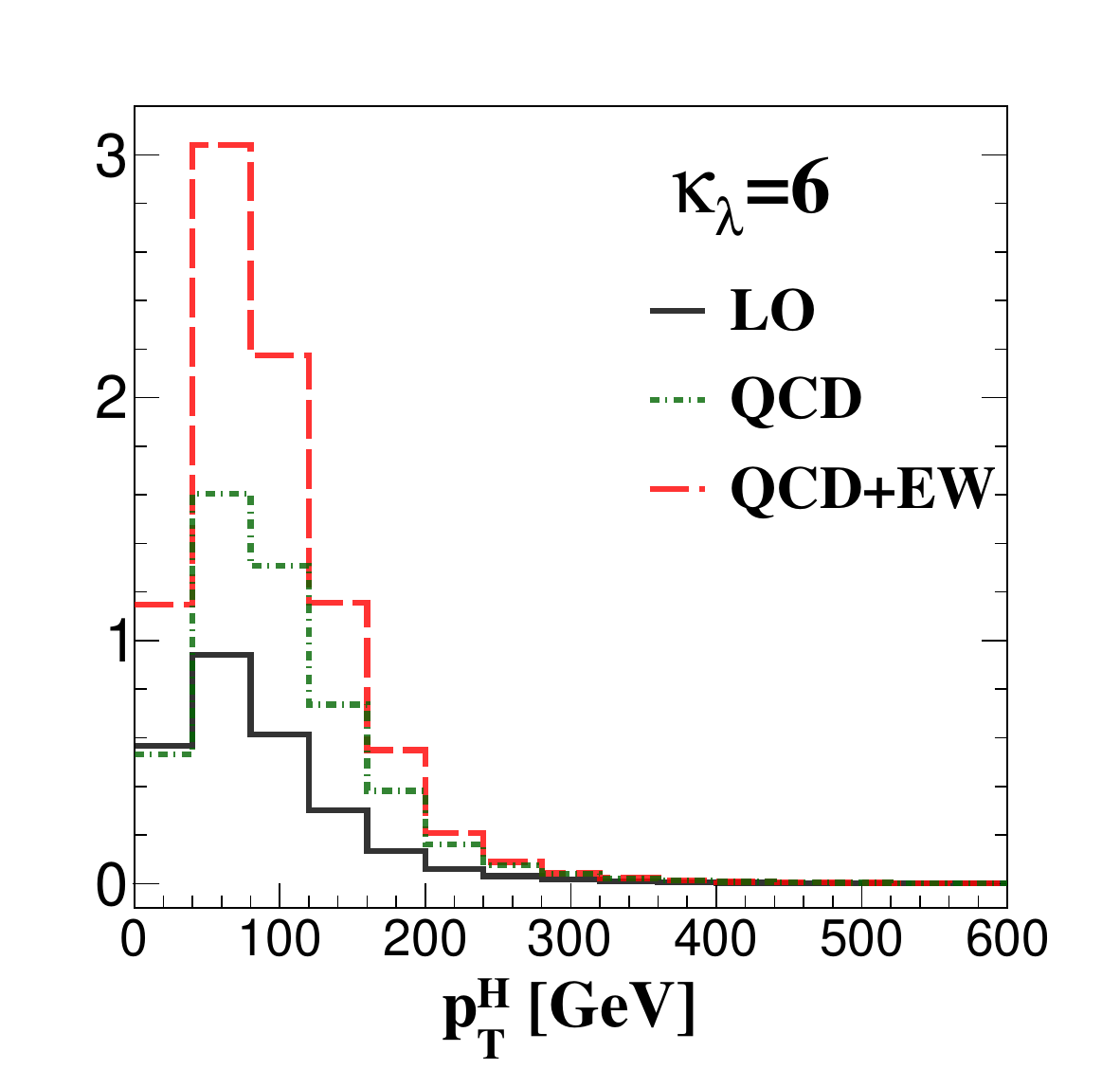}	
\vspace{2mm} 
\caption{Kinematic distributions for double-Higgs production in the combined ggF and VBF channels at the LHC with $\sqrt{s} = 14 \, {\rm TeV}$ for three different $\kappa_\lambda$ values, assuming $\kappa_\lambda = \kappa_3 = \kappa_4$. The black lines show the LO predictions, the green dash-dotted lines include QCD corrections, and the red dashed lines incorporate both QCD and EW~corrections. The upper panels display $m_{HH}$ spectra, while the lower panels show~$p_{T}^{H}$ distributions.} 
\label{fig:dist-kappa}
\end{figure}

In the left plot of Figure~\ref{fig:kappa}, we show the cross sections for double-Higgs production in the combined~ggF and VBF channels at the LHC with $\sqrt{s} = 14 \, \mathrm{TeV}$ as a function of $\kappa_\lambda = \kappa_3 = \kappa_4$ at different perturbative orders in the HEFT. Specifically, the black line corresponds to the LO predictions given in Eqs.~(\ref{eq:ggFLO}) and~(\ref{eq:VBF}), while the green line shows the higher-order QCD predictions from Eqs.~(\ref{eq:ggFNLO}) and~(\ref{eq:VBF}). 
The~red line additionally includes the NLO~EW~effects of Eqs.~(\ref{eq:typeI}), (\ref{eq:EWIIggF}), and~(\ref{eq:EWIIVBF}), assuming that the EW~and QCD corrections are factorizable. We~first note that, for the considered range of $\kappa_\lambda$, the VBF process contributes between $6\%$ and $19\%$ of the total double-Higgs production cross section. One furthermore observes that for $\kappa_\lambda \gtrsim 3.5$, the NLO~EW~effects are as important as, or even more important than, the QCD corrections. This is phenomenologically relevant, as incorporating NLO~EW~effects into the experimental analyses of double-Higgs production would allow the upper limits of the two-sided bounds in~Eqs.~(\ref{eq:kappa3bounds}) to be tightened. We add that, while the results shown assume $\kappa_\lambda = \kappa_3 = \kappa_4$, the plot remains essentially unchanged if $\kappa_4$ is allowed to vary within the range $-10 \lesssim \kappa_4 \lesssim 10$. 

The latter feature is also reflected by the red contour in the right panel of Figure~\ref{fig:kappa}, which shows the $68\%$~CL region derived from the HEFT calculation of QCD and EW~corrections to double-Higgs production presented in this section, applied to~LHC~Run~2 data. As in the two panels of~Figure~\ref{fig:planes}, it is evident that even after including NLO~EW~corrections involving $\kappa_3$ and $\kappa_4$, double-Higgs production shows only a weak dependence on the coupling modifier~$\kappa_4$, reflecting the loop suppression of the contribution arising from the Higgs quartic self-coupling. For~comparison, the right panel of~Figure~\ref{fig:kappa} also displays the $68\%$~CL region from triple-Higgs production, as obtained by the ATLAS collaboration in~Ref.~\cite{ATLAS:2024xcs}, shown as a black contour. The~constraints on~$\kappa_4$ from triple-Higgs production are notably stronger than those from double-Higgs production, due to the fact that in $gg \to HHH$ the Higgs quartic self-coupling already appears at LO in perturbation~theory.  Finally, note that, in contrast to~Figure~\ref{fig:planes}, the constraints on $\kappa_4$ from double-Higgs production shown on the right of~Figure~\ref{fig:kappa} exhibit a flat direction. This arises because the HEFT results include only terms linear in $\kappa_4$, as~evident from Eqs.~(\ref{eq:muhhsHEFT}). By contrast, the SMEFT expressions in Eqs.~(\ref{eq:muhhs}) contain~$\kappa_4^2$ contributions, which break this flat direction. It is important to note that this difference is not related to the choice of SMEFT or HEFT, but rather to whether the square of the one-loop BSM contribution proportional to $\kappa_4$ is included.

The NLO~EW~corrections discussed in this section do not only modify the total double-Higgs production cross sections but also leave interesting imprints in the kinematic distributions. To~illustrate this point, Figure~\ref{fig:dist-kappa} shows the $m_{HH}$ spectra as well as the distributions of the transverse momentum of the leading Higgs boson ($p_{T}^{H}$). The displayed kinematic distributions correspond to the combined ggF and VBF predictions at the LHC with $\sqrt{s} = 14 \, {\rm TeV}$ for three representative values of $\kappa_\lambda$, assuming again~$\kappa_\lambda = \kappa_3 = \kappa_4$. In the SM, i.e., for $\kappa_\lambda = 1$, the $m_{HH}$ spectrum exhibits a peak around $400 \, {\rm GeV}$. In general, this peak shifts towards smaller $m_{HH}$ values as $\kappa_\lambda$ increases. However, larger values of $\kappa_\lambda$ can also induce a dip in the spectrum, as illustrated in the upper middle panel for $\kappa_\lambda = 3$. It is further evident that the NLO~EW~corrections grow in importance with increasing $\kappa_\lambda$ and can eventually exceed the size of the QCD corrections around the peak. Furthermore, the NLO~EW~corrections do not enhance the~$m_{HH}$ distribution uniformly, as their impact is more pronounced near the peak. The $p_T^H$ distributions exhibit similar behavior when varying $\kappa_\lambda$. Given the sizable and non-trivial impact of the NLO~EW~corrections on the kinematic distributions for $\kappa_{\lambda} \gtrsim 3$, including these effects is expected to improve the sensitivity to $\kappa_3$ and $\kappa_4$ in an actual LHC data analysis, which necessarily relies on an exclusive phase-space~selection.

\section{Conclusions}
\label{sec:conclusions}

Compared with the Higgs couplings to EW~gauge bosons or third-generation fermions, the Higgs self-couplings remain only weakly constrained at LHC~Run~2. This leaves substantial room for BSM~physics that predominantly modifies the shape of the Higgs potential. In~fact, theoretical arguments supported by explicit UV-complete model calculations show that Higgs self-coupling deviations exceeding $200\%$ are still compatible with existing single-Higgs measurements, without requiring fine-tuning of UV~parameters~\cite{Durieux:2022hbu}. Probing the Higgs self-couplings at the HL-LHC and future colliders via multi-Higgs production therefore explores largely uncharted model and parameter space and provides genuine discovery potential, even if single-Higgs observables show no anomalies.

Fully exploiting the discovery potential of multi-Higgs production channels requires including higher-order perturbative corrections induced by both QCD and EW~interactions in the corresponding theoretical predictions. For double-Higgs production, such calculations have been performed in Refs.~\cite{Bizon:2018syu,Borowka:2018pxx,Li:2024iio} using two consistent EFT approaches. The first two works~\cite{Bizon:2018syu,Borowka:2018pxx} employed the SMEFT to compute two-loop EW~corrections involving the coupling modifiers $\kappa_3$ and $\kappa_4$, while the third~\cite{Li:2024iio} carried out the calculation within the HEFT framework. We have summarized the main assumptions and results of the SMEFT and HEFT calculations in~Sections~\ref{sec:SMEFT}~and~\ref{sec:HEFT}, respectively. From these discussions, it is clear that the two calculations differ in several conceptual and technical aspects, rendering a direct analytic comparison of the corresponding master formulas in Eqs.~(\ref{eq:muhhs}) and (\ref{eq:muhhsHEFT}) beyond the scope of this~note.

For clarity, let us briefly highlight these differences. First, the SMEFT and HEFT are distinct~EFTs, with the Higgs being part of an $SU(2)_L$ doublet in the SMEFT and treated as a singlet in the HEFT. This distinction implies, for example, that the Wilson coefficients $C_{2n}$ of the $(\phi^\dagger \phi)^n$ operators in the SMEFT exhibit a different renormalization-group~(RG) evolution compared to the coupling modifiers $\kappa_n$ of the $H^n$ operators in HEFT with $n \geq 3$. The UV structure of the two EFTs is hence not the same. Second, the SMEFT calculation reviewed in~Section~\ref{sec:SMEFT} includes contributions from the Higgs trilinear, quartic, and quintic self-couplings, using~Eq.~(\ref{eq:kappa5}) to express $\kappa_5$ in terms of $\kappa_3$ and~$\kappa_4$. The HEFT computations instead include $\kappa_3$ and~$\kappa_4$~while setting $\kappa_5 = 0$. Including corrections from the Higgs quintic self-coupling in the HEFT framework is straightforward, but it would modify the formulas such as Eqs.~(\ref{eq:kappa3ct}), (\ref{eq:EWIIggF}), and (\ref{eq:EWIIVBF}) presented in Section~\ref{sec:HEFT}. Third, the two NLO~EW~calculations retain different orders in the expansion in $\kappa_3$ and $\kappa_4$. In the SMEFT computation of~Section~\ref{sec:SMEFT}, the square of the two-loop EW~diagrams is included, so the results in~Eqs.~(\ref{eq:muhhs}) feature a $\kappa_4^2$ dependence. In~contrast, the HEFT calculation detailed in~Section~\ref{sec:HEFT} considers only the interference between the two-loop EW~diagrams and the one-loop diagrams, yielding a linear $\kappa_4$ dependence in~Eqs.~(\ref{eq:EWIIggF}). The~same choice is adopted in the SMEFT calculation of Ref.~\cite{Borowka:2018pxx}. However, only~for large values of $\kappa_4$ --- i.e.,~parameters outside the region where the red and yellow constraints overlap in the plot on the left-hand-side in~Figure~\ref{fig:planes} --- does~this computation lead to predictions that differ from the parametrization presented in~Eqs.~(\ref{eq:muhhs}), which is based on the results obtained in Refs.~\cite{Bizon:2018syu,Bizon:2024juq}. In addition, the~HEFT calculation of $gg \to HH$ incorporates terms of~${\cal O} (\kappa_3^3)$ and~${\cal O} (\kappa_3^4)$, as well as the corrections in Eqs.~(\ref{eq:typeI}), which are not included in the SMEFT results. Fourth and finally, the phenomenological HEFT results also account for higher-order QCD and EW~effects in the VBF double-Higgs production process, whereas the corresponding SMEFT results include only the~ggF~channel.

Despite these differences, the phenomenological implications of the two calculations are in good agreement. This is illustrated by the constraints in the $\kappa_3\hspace{0.25mm}$--$\hspace{0.25mm}\kappa_4$ plane derived from LHC~Run~2 double-Higgs production data, shown on the left in~Figure~\ref{fig:planes} and on the right in~Figure~\ref{fig:kappa}, respectively. A more detailed comparison of the SMEFT and HEFT results is presented in Appendix~\ref{app:SMEFTHEFT}. In fact, in the parameter region allowed by perturbative unitarity, the two NLO~EW~computations of double-Higgs production discussed in this note yield numerically similar constraints. This agreement highlights the model-independent complementarity of double- and triple-Higgs production in probing the Higgs potential, with double-Higgs production providing stronger constraints on $\kappa_3$ and triple-Higgs production on $\kappa_4$. With~the~HL-LHC on the horizon and the prospect of future high-energy colliders such as the FCC becoming increasingly tangible, there is a strong motivation to further refine the theoretical description of double-Higgs production. In particular, combining the two NLO~EW~calculations presented here in a common, systematic framework would facilitate a more robust interpretation of multi-Higgs production measurements at the HL-LHC and future colliders. This task, however, requires significant theoretical work and is therefore left for future research endeavors.
 
\section*{Acknowledgements}

The Feynman diagrams in this note were drawn using \texttt{TikZ-Feynman}~\cite{Ellis:2016jkw}. This work was carried out within the LHC Higgs Working Group as a contribution to the CERN Report~5. We~thank the members of the working group for valuable discussions, and A.~Carvalho Antunes de Oliveira, A.~Karlberg, G.~Landsberg, F.~Monti, L.~Scyboz and D.~Stolarski for their helpful comments, which allowed us to improve the manuscript. The research of J.-L.~Ding, H.~T.~Li, Z.~G.~Si, J.~Wang, X.~Zhang, and D.~Zhao was supported by the National Natural Science Foundation of China under Grants No.~12321005 and No.~12375076. D.~Pagani acknowledges financial support by the MUR through the PRIN2022 Grant 2022EZ3S3F, funded by the European Union --- NextGenerationEU.

\begin{appendix}

\section{Higgs self-couplings and SMEFT operator truncation}
\label{app:SMEFT}

In this appendix, we show that restricting the SMEFT Lagrangian to pure Higgs operators of the form $(\phi^\dagger \phi)^n$, as in~Eq.~(\ref{eq:LSMEFT}), is well motivated both phenomenologically and in explicit UV-complete models. Our discussion builds on~\cite{FCC26UH}. The key point is that current constraints from Higgs measurements and EW precision observables~(EWPOs) imply a hierarchy among SMEFT Wilson coefficients, such that deviations of the Higgs trilinear self-coupling are dominantly controlled by the operator $(\phi^\dagger \phi)^3$.

Before discussing the SMEFT operators that directly modify the Higgs self-couplings at tree level, it is helpful to first briefly review other effective interactions that can also influence double-Higgs production. Two widely studied dimension-six operators are
\beq \label{eq:2outof2499}
Q_{\phi G} = (\phi^\dagger \phi) \hspace{0.5mm} G_{\mu \nu}^a G^{a, \mu \nu} \,, \qquad
Q_{\phi t} = (\phi^\dagger \phi) \hspace{0.5mm} \bar q t \widetilde{\phi} \,,
\eeq
where $G_{\mu \nu}^a$ is the QCD field strength tensor, $q$ denotes the left-handed third-generation quark doublet, and we use the shorthand $\widetilde{\phi}_i = \epsilon_{ij} \left(\phi^j\right)^\ast$, with $\epsilon_{ij}$ the fully antisymmetric Levi-Civita tensor normalized as $\epsilon_{12} = +1$. The first operator induces an effective coupling between the Higgs and gluons, while the second modifies the top-quark Yukawa coupling relative to the~SM. Consequently, the operators in~Eqs.~(\ref{eq:2outof2499}) are also constrained by single-Higgs production and decay. Since the leading single-Higgs production processes are measured with an accuracy of roughly~$10\%$, the Wilson coefficients of $Q_{\phi G}$ and $Q_{\phi t}$ are constrained to be relatively small. These operators contribute to single-Higgs production already at LO, which tightly bounds their size. As a result, their effects on multi-Higgs production are subleading. In the case of triple-Higgs production, this has been demonstrated recently in~\cite{Panizzi:2025sya}. Therefore, in analyses such as those presented in~Section~\ref{sec:SMEFT}, the contributions of Eqs.~(\ref{eq:2outof2499}) and other operators affecting single-Higgs production and decay can be neglected to first approximation.

We now turn to the tree-level matching condition that relates the relevant Wilson coefficients of the dimension-six SMEFT operators to the deviation $\delta \kappa_3$ of the Higgs trilinear self-coupling:
\beq \label{eq:deltakappa3} 
\delta \kappa_3 \simeq -\frac{2 \hspace{0.25mm} v^4}{m_h^2} \, \frac{C_6}{\Lambda^2} + \frac{v^2}{\Lambda^2} \, \left( C_{\phi \Box} - \frac{1}{4} C_{\phi D} \right) \,.
\eeq
Here, $C_6$ is the Wilson coefficient of the dimension-six operator introduced in~Eq.~(\ref{eq:dim67}), while $C_{\phi \Box}$ and $C_{\phi D}$ are the dimensionless Wilson coefficients of
\beq \label{eq:dim76}
Q_{\phi \Box} = (\phi^\dagger \phi) \, \Box \, (\phi^\dagger \phi) \,, \qquad Q_{\phi D} = (\phi^\dagger D_\mu \phi)^\ast (\phi^\dagger D^\mu \phi) \,,
\eeq
with $\Box = \partial_\mu \partial^\mu$ denoting the d'Alembert operator. Note that in~(\ref{eq:deltakappa3}) possible SMEFT corrections to the Fermi constant $G_F$ are neglected, as they are tightly constrained phenomenologically. The Higgs mass and VEV are treated as EW input parameters, so any SMEFT effects are absorbed into their definitions and do not appear explicitly as corrections from the Wilson~coefficients.

The first important observation is that the operators in~Eq.~(\ref{eq:dim76}) also induce tree-level contributions to other observables. In particular, the modification $\delta \kappa_V$ of the Higgs coupling to EW gauge bosons and the Peskin-Takeuchi $\hat T$ parameter~\cite{Peskin:1990zt} --- an important EWPO~that measures custodial-symmetry breaking --- are given by
\beq \label{eq:deltakappaVT}
\delta \kappa_V \simeq \frac{v^2}{\Lambda^2} \, C_{\phi \Box} \,, \qquad 
\hat T \simeq -\frac{v^2}{2 \hspace{0.25mm} \Lambda^2} \, C_{\phi D} \,.
\eeq
In contrast to $\delta \kappa_3$, which is currently only weakly constrained~(\ref{eq:kappa3bounds}), the parameters $\delta \kappa_V$ and $\hat T$ are subject to much stronger experimental bounds
\beq \label{eq:dkVhatTexp}
|\delta \kappa_V| \lesssim 10\% \,, \qquad |\hat T| \lesssim 0.1\% \,,
\eeq
where the limits are taken from~\cite{ATLAS:2025qxq,CMS:2026nce} and~\cite{ALEPH:2005ab,ParticleDataGroup:2024cfk}, respectively.

Eqs.~(\ref{eq:deltakappa3}) and~(\ref{eq:deltakappaVT}) can now be used to derive model-independent relations among the Wilson coefficients
\beq \label{eq:LHCLEPhierarchy} 
\frac{|C_6|}{|C_{\phi \Box}|} \simeq \frac{m_h^2}{2 \hspace{0.25mm} v^2} \frac{|\delta \kappa_3|}{|\delta \kappa_V|} \simeq 7 \,, \qquad 
\frac{|C_{\phi \Box}|}{|C_{\phi D}|} \simeq \frac{|\delta \kappa_V|}{2 \hspace{0.25mm} |\hat T|} \simeq 50 \,,
\eeq
where for $\delta \kappa_3$ only the contribution of $C_6$ to~Eq.~(\ref{eq:deltakappa3}) is retained. The numerical values are obtained by adopting the maximal allowed values of $|\delta \kappa_3|$, $|\delta \kappa_V|$, and $|\hat T|$ from Eqs.~(\ref{eq:kappa3bounds}) and~(\ref{eq:dkVhatTexp}). The results in~Eq.~(\ref{eq:LHCLEPhierarchy}) imply that, to satisfy constraints from Higgs measurements and EWPOs, the Wilson coefficients must exhibit a hierarchical structure:
\beq \label{eq:hierarchy}
|C_6| > |C_{\phi \Box}| \gg |C_{\phi D}| \,.
\eeq
Kitchen sink BSM scenarios predicting $|C_6| \simeq |C_{\phi \Box}| \simeq |C_{\phi D}| \simeq {\cal O} (1)$ are therefore ruled out. Note that the first relation in~Eq.~(\ref{eq:LHCLEPhierarchy}) also shows that the projected HL-LHC limit of $50\%$ on $\delta \kappa_3$ provides roughly a factor of three less sensitivity than the anticipated $2\%$ HL-LHC measurement of $\delta \kappa_V$~\cite{ATL-PHYS-PUB-2025-018}.

Realizing the hierarchy~(\ref{eq:hierarchy}) within concrete UV-complete scenarios requires some model-building effort. A canonical example is the custodial quadruplet model~\cite{Durieux:2022hbu}, which naturally allows a large ratio of the Higgs trilinear self-coupling modification relative to other Higgs coupling deviations, while keeping contributions to EWPOs sufficiently suppressed. In this model, the relevant Wilson coefficients are calculable:
\beq \label{eq:CQmatching} 
\frac{C_6}{\Lambda^2} \simeq \frac{2 \hspace{0.25mm} \lambda_\Theta^2}{3 \hspace{0.25mm} M_\Theta^2} \,, \qquad \frac{C_{\phi \Box}}{\Lambda^2} \simeq \frac{\lambda_\Theta^2}{4 \pi^2 \hspace{0.25mm} M_\Theta^2} \,, \qquad \frac{C_{\phi D}}{\Lambda^2} \simeq -\frac{5 \hspace{0.25mm} g_1^2}{12 \pi^2} \, \ln \left( \frac{M_\Theta}{m_Z} \right) \, \frac{C_{\phi \Box}}{\Lambda^2} \,.
\eeq
Here, $\lambda_\Theta$ denotes the coupling of the operator $H^3 \hspace{0.25mm} \Theta$, with $\Theta$ the custodial quadruplet, and $M_\Theta$ its mass. Eq.~(\ref{eq:CQmatching}) shows that the hierarchy~(\ref{eq:hierarchy}) is naturally realized because $C_6$, $C_{\phi \Box}$, and~$C_{\phi D}$ appear at different perturbative orders: $C_6$ at tree level, $C_{\phi \Box}$ at one loop, and $C_{\phi D}$ at two loops. Note that the expression for $C_{\phi D}$ includes only the leading-logarithmic correction from the one-loop RG evolution of $Q_{\phi \Box}$ into $Q_{\phi D}$~\cite{Alonso:2013hga}, which provides the leading contribution for sufficiently large $M_\Theta$. The loop suppression of $C_{\phi \Box}$ and $C_{\phi D}$ ensures that, for $\delta \kappa_3$ as given in~Eq.~(\ref{eq:deltakappa3}), contributions from the pure Higgs operator $Q_6$ dominate, so that the relevant physics is well captured by~Eq.~(\ref{eq:LSMEFT}).

The same argument applies to other BSM frameworks that can primarily modify the Higgs potential. Notable examples include the real singlet model with a $Z_2$ symmetry~\cite{Haisch:2020ahr,Maura:2024zxz} and the aligned, approximately custodial-symmetric two-Higgs-doublet model with TeV-scale pseudoscalar and charged Higgs masses~\cite{Bahl:2022jnx}. In these models, both $C_6$ and $C_{\phi \Box}$ are generated at one loop, but the matching condition for $C_6$ depends on higher powers of the quartic scalar couplings than that of $C_{\phi \Box}$. Universal contributions to the Wilson coefficient $C_{\phi D}$ are generated at the two-loop level, as in the custodial quadruplet model. Consequently, in the limit of large quartic scalar couplings a hierarchy of the form~(\ref{eq:hierarchy}) emerges, and the correction to the Higgs trilinear self-coupling~(\ref{eq:deltakappa3}) is fully dominated by $Q_6$. Hence, restricting the SMEFT Lagrangian to operators of the form $(\phi^\dagger \phi)^n$, like in~Eq.~(\ref{eq:LSMEFT}), to effectively describe these BSM models is again justified.

We have demonstrated that UV-complete models can realize the phenomenologically required hierarchy~(\ref{eq:hierarchy}), in which case truncating the SMEFT Lagrangian as in~Eq.~(\ref{eq:LSMEFT}) provides a reliable approximation. Since the matching between the BSM model and the SMEFT is performed at the high scale set by the masses of the new states, whereas constraints such as those in Figure~\ref{fig:planes} involve Wilson coefficients evaluated at the EW scale, one must verify that the SMEFT truncation remains valid under the RG evolution. This is indeed the case for the operator~$Q_6$, which at the one-loop level only mixes into itself~\cite{Jenkins:2013zja,Jenkins:2013wua,Alonso:2013hga} and at two loops only into the dimension-six top-quark Yukawa-type operator~\cite{Gorbahn:2016uoy,Born:2026xkr}. In a UV-complete model where~$Q_6$ provides the dominant dimension-six SMEFT deformation, this dominance is thus, to very good approximation, stable under the RG evolution. 

\section{Numerical comparison of SMEFT and HEFT predictions}
\label{app:SMEFTHEFT}

\begin{figure}[t!]
\begin{center}
\includegraphics[height=0.45\textwidth]{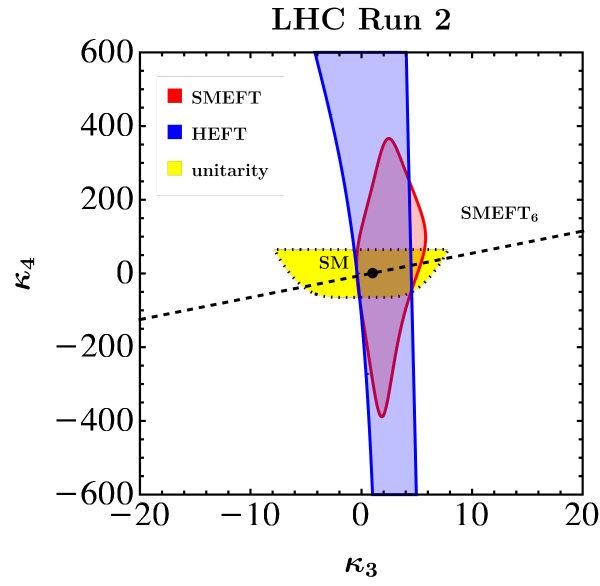} \qquad 
\includegraphics[height=0.45\textwidth]{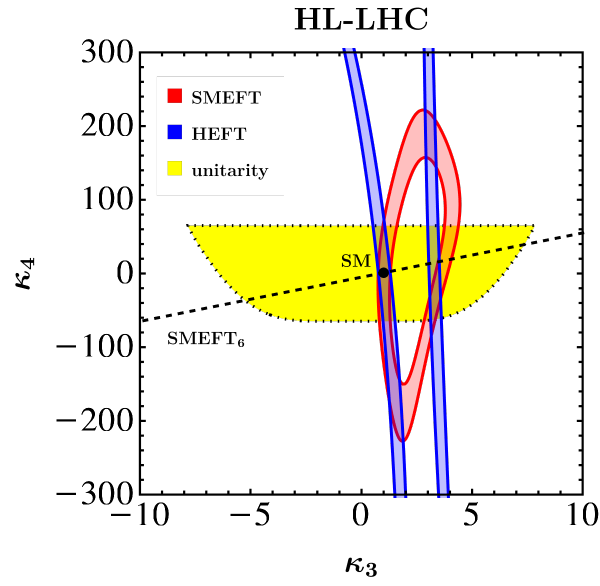}
\end{center}
\vspace{-4mm} 
\caption{\label{fig:SMEFTHEFTcomparison} Constraints in the $\kappa_3\hspace{0.25mm}$--$\hspace{0.25mm}\kappa_4$~plane similar to that shown in Figure~\ref{fig:planes}. The red and blue contours indicate the constraints derived from double-Higgs production in the SMEFT and HEFT, respectively. These constraints are obtained using Eqs.~(\ref{eq:muhhs}) and (\ref{eq:muhhsHEFT}).}
\end{figure}

This appendix provides a more detailed numerical comparison of the SMEFT and HEFT results for double-Higgs production discussed in the main text. Specifically, we compare the phenomenological constraints on $\kappa_3$ and $\kappa_4$ that follow from Eqs.~(\ref{eq:muhhs}) and (\ref{eq:muhhsHEFT}), respectively, when applied to existing LHC~Run~2 data and to hypothetical HL-LHC measurements with an assumed integrated luminosity of $3 \, {\rm ab}^{-1}$. The results are shown in the two panels of~Figure~\ref{fig:SMEFTHEFTcomparison}, with red and blue contours indicating the constraints derived from double-Higgs production in the SMEFT and HEFT, respectively. These constraints are obtained by requiring $\mu_{2h}^{\textrm{LHC~Run~2}} < 2.9$~\cite{ATLAS:2024ish} and $0.77 < \mu_{2h}^{\textrm{HL-LHC}} < 1.23$~\cite{ATL-PHYS-PUB-2022-005} on the double-Higgs production signal strength at LHC~Run~2 and HL-LHC.

For the LHC~Run~2 constraints shown on the left in~Figure~\ref{fig:SMEFTHEFTcomparison}, one observes that, within the parameter region allowed by perturbative unitarity --- indicated by the yellow areas outlined with black dotted lines --- the two NLO EW computations of double-Higgs production discussed in this note yield numerically very similar constraints in the $\kappa_3\hspace{0.25mm}$--$\hspace{0.25mm}\kappa_4$~plane. The same holds for the region near $\kappa_4 - 1 = 6 \, (\kappa_3 - 1)$, represented by the black dashed line, which corresponds to the relation between $\kappa_3$ and $\kappa_4$ that arises in the SMEFT at the level of dimension-six operators. For $|\kappa_4| \gtrsim 65$, however, the two constraints begin to deviate. In particular, the HEFT constraint exhibits a flat direction, which arises because the HEFT results include only terms linear in~$\kappa_4$, as evident from Eqs.~(\ref{eq:muhhsHEFT}). By contrast, the SMEFT expressions in Eqs.~(\ref{eq:muhhs}) contain~$\kappa_4^2$ contributions, which break this flat direction. Technically, this difference arises because the SMEFT computation in~Section~\ref{sec:SMEFT} includes the square of the two-loop BSM diagrams, while the HEFT calculation in~Section~\ref{sec:HEFT} considers only their interference with the one-loop SM diagrams.

As shown in the right panel of~Figure~\ref{fig:SMEFTHEFTcomparison}, the impact of $\kappa_4^2$ terms on the resulting constraints is even more pronounced in the case that future double-Higgs production measurements yield a two-sided limit on the signal strength, as expected at the HL-LHC. Since the $\kappa_4^2$ terms are absent in the HEFT results reported in~Eqs.~(\ref{eq:muhhsHEFT}), the corresponding constraints now form two bands that are nearly parallel to the $\kappa_4$ axis: one passes through the SM point, while the other includes the BSM point $\{\kappa_3, \kappa_4\} \simeq \{3.5, 16\}$. Due to the presence of $\kappa_4^2$ terms in~Eqs.~(\ref{eq:muhhs}) the SMEFT constraint instead forms a elongated ellipsoidal band. While the global overall agreement between the SMEFT and HEFT constraints is therefore not particularly strong, one observes that for solutions near the SM point and within the theoretical limits imposed by perturbative unitarity, the SMEFT and HEFT constraints match quite well. A similar behavior is observed in the vicinity of the BSM point $\{\kappa_3, \kappa_4\} \simeq \{3.5, 16\}$, which lies on the line $\kappa_4 - 1 = 6 \, (\kappa_3 - 1)$, corresponding to the parameter space where $Q_6$ in~Eqs.~(\ref{eq:LSMEFT}) is the only SMEFT operator with a non-zero Wilson coefficient. Here, the agreement between the two constraints is somewhat worse, due to the presence of $\kappa_3^3$ and $\kappa_3^4$ terms in Eqs.~(\ref{eq:muhhsHEFT}) that are absent in Eqs.~(\ref{eq:muhhs}). Including these higher-order terms in Eqs.~(\ref{eq:muhhs}) would noticeably improve the agreement of the constraints near this BSM point. Note that, from a model-building perspective, the parameter space around and between these two points appears particularly relevant, given the hierarchy in~Eq.~(\ref{eq:hierarchy}) and the fact that, in BSM models such as the custodial quadruplet model discussed in~Appendix~\ref{app:SMEFT} that realizes it, one generically expects $|C_8| \simeq |C_6|$. Put differently, BSM~models that lead to $|\kappa_4| \gg |\kappa_3|$ appear notoriously difficult to construct --- we are not aware of any model that realizes this --- implying that the differences between the SMEFT and HEFT constraints shown in~Figure~\ref{fig:SMEFTHEFTcomparison} are likely of limited phenomenological relevance when confronted with explicit perturbative UV-complete models. We finally note that incorporating kinematic information in double-Higgs production at the HL-LHC is expected to eliminate the solutions near $\{\kappa_3, \kappa_4\} \simeq \{3.5, 16\}$, leaving only SM-like solutions as viable in the $\kappa_3\hspace{0.25mm}$--$\hspace{0.25mm}\kappa_4$~plane, where the SMEFT and HEFT predictions show rather good agreement.

\end{appendix}



\end{document}